\documentclass[aps,pra, two column,10pt,superscriptaddress,nofootinbib, notitlepage]{revtex4-2}

\newcommand\identity{1\kern-0.25cm\text{l}}
\usepackage{etex}
\usepackage{physics}
\usepackage{amsmath,amssymb,amsthm,amstext,mathrsfs}
\usepackage[colorlinks=true,citecolor=blue,urlcolor=blue]{hyperref}
\usepackage[T1]{fontenc}
\usepackage{comment}
\usepackage{eufrak}
\usepackage{enumerate}
\usepackage{times,txfonts}
\usepackage{braket}
\usepackage{svg}
\usepackage{color}
\usepackage{natbib}
\usepackage{amsmath,blkarray}
\usepackage{mathtools}
\usepackage{latexsym}
\usepackage{tabularx, booktabs}
\usepackage{epstopdf}
\usepackage{lipsum}
\usepackage{float}
\usepackage{amsfonts}
\usepackage{subcaption}
\usepackage{color,soul}
\usepackage{tikz}
\usepackage{amssymb,mathrsfs}
\usetikzlibrary{positioning,arrows.meta,calc}

\newcommand{\be}{\begin{equation}}
	\newcommand{\ee}{\end{equation}}
\newcommand{\ba}{\begin{eqnarray}}
	\newcommand{\ea}{\end{eqnarray}}

\definecolor{ss}{RGB}{250,80,220}
\begin{document}

\title{Semi-Device-Independent Certification of multi-qubit Entangled Basis Measurement}
\author{Rajat Pandey}
\email{rajatpandey2062002@gmail.com}
\author{A.K.Pan}
\email{akp@phy.iith.ac.in}
\affiliation{Indian Institute of Technology Hyderabad, Kandi, Sangareddy, Telangana 502285, India }

\begin{abstract}
The entangled basis measurement plays a pivotal role in fundamental information processing tasks, such as, dense coding, quantum teleportation and entanglement swapping. Consequently, certifying such measurements in a device-independent manner remains an important task. In this work, we present a semi-device-independent(SDI) protocol in prepare-measure (PM) scenario for certifying a class of $m$-qubit entangled basis measurement. Specifically, we consider tri-partite PM scenario featuring two senders and one receiver where each sender sends arbitrary $2^{m}$ number of inputs and the receiver perform a  $2^m$ outcome  measurement. We adopt a classical causal-model framework to demonstrate the nonclassicality of the quantum measurements.  We introduce nonlinear  witnesses in classical causal models and demonstrate that the optimal quantum value exceeds the classical bounds. Further, we demonstrate that the optimal quantum value of the witness SDI certify the entangled basis measurement and the prepared states of the two senders.  
\end{abstract}

\maketitle

\section{Introduction}
It is well established that not all correlations predicted by quantum theory are compatible with the theories that respect local realism. This is the famous no-go theorem put forward by Bell ~\cite{J.S.BellPhysicsPhysiqueFizika.1.195}.  This seminal result has since given rise to extensive developments in foundational studies as well as practical applications in quantum information science (see, reviews ~\cite{bell_nonlocality_2014_RevModPhys.86.419,Quantum_cryptography_2002_RevModPhys.74.145,Tavakoli_2022}). One of the remarkable features of Bell's theorem is that it enables DI self-testing \cite{mayers2004selftestingquantumapparatus,Supic2020selftestingof}, which is the strongest form of certification of quantum resources solely from observed input-output statistics. In such protocols, the internal mechanism of the devices remains uncharacterized, and the dimensionality of the underlying quantum systems remains unknown. From a theoretical perspective,  self-testing probes whether an underlying quantum resource can generate unique realizations up to some local isometry. 
  
The notion of self-testing was first introduced by Mayer and Yao \cite{mayers2004selftestingquantumapparatus}. Since then a flurry of ingenious protocols have been developed. For example, the DI self-testing of pure non-maximally entangled two-qubit states \cite{Acin2012,  Coladangelo2017, Wooltron2022, Rai2022, Bamps2015}, Pauli observables \cite{Bowles2018, Pan2021}, quantum instruments \cite{Wagner2020,mohan_2019}, quantum memory \cite{SekatskiPrl2018, SekatskiPrl2023}, unsharp measurements \cite{Roy2023, Paul2024,Gomez2016}, unitary gates\cite{Sarkar_unitary_gate2026)}   and so on. This approach has been extended to multipartite systems \cite{Wu2014,Panwar2023, ritesh_GHz_6ct-vmnx,McKague10.1007/978-3-642-54429-3_7} and multi-outcome measurement scenario \cite{Sarkar2021}. Extending further to the network scenarios featuring multiple sources, the  self-testing of all entangled states \cite{Supic2023QuantumNetworksSelfTest}, commuting observables \cite{A.K.pan_2022_PhysRevA.105.032216,sneha_et_al._2023_PhysRevA.108.062607}, and complex quantum theory \cite{Renou2021} have been reported. For a recent review, we refer to \cite{Supic2020selftestingof}.   

However, the practical implementation of a loophole-free Bell test offers stringent challenges. A potential alternative is SDI certification in PM scenarios. In this direction, several works have explored certification protocols for various quantum resources and operations, including quantum states and measurements \cite{Navascués_2023_PM_PhysRevLett.131.250802, Tavakoli_2018,supic_2020_PM,pauwels_2022_PM_PhysRevLett.129.250504,_Tavakoli_et.al_QRAC_PhysRevLett.114.170502}, unitary operations \cite{Paul_2025}, dimension witnesses \cite{vertesi_et.al_dimension_witness_2013_PhysRevLett.110.150501}, unsharpness parameters \cite{Sumit_2021PhysRevA.104.062214,Wei_2021,Abhyoudai_et.al_PhysRevA.107.012411} and quantum key distribution \cite{Brunner_et.al_SDIQKD_2011_PhysRevA.84.010302} and so on
In a general PM scenario, self-testing of the underlying physical system relies on additional assumptions, including knowledge of the Hilbert space dimension \cite{seft_testing_in_prepare_measure_PhysRevA.98.062307}, energy constraints \cite{VanHimbeeck2017semidevice}, fidelity and minimum-overlap constraints \cite{Tavakoli_fidelity_PhysRevA.100.042108,Burnner_etal_overlap_PhysRevApplied.7.054018}, and restrictions on the information carried by the physical system \cite{Tavakoli2020informationally,Chaves_2015_PhysRevLett.115.110501,Tavakoli2022informationally,ritesh_Singh_2025,Chailloux_2016}.

 Entangled basis measurements play a pivotal role in quantum information theory, in particular in fundamental protocols such as entanglement swapping \cite{entanglement_swapping_1993PhysRevLett.71.4287}, dense coding \cite{Dense_coding_1992}, and quantum teleportation \cite{Teleportation_95}. This makes certification of entangled basis measurements an important task. An attempt was made in \cite{post_selection_PhysRevLett.107.050502} in the bilocal network scenario through the post-selection of central party measurement. The approach in  \cite{sdi_certification_2011_PhysRevA.83.062112,sdi_exp_certification_PhysRevLett.113.080405} uses entanglement witnesses to certify the measurement, assuming knowledge of the underlying quantum system. However,  those approaches certify the presence of entanglement in the measurement basis but do not characterize a particular entangled measurement. In the network scenario, self-testing protocols have been developed for entangled basis measurements \cite{Renou_2018,Sarkar2026}, but requires pre-shared entanglement between the parties.
 
In this work, we develop a SDI protocol to certify multi-qubit entangled basis measurements in PM scenario. A standard PM scenario commonly involves two parties, one sender and one receiver. We consider a PM scenario involving two senders and one receiver where the senders  prepare the physical systems and send to the receiver. The receiver then performs a joint measurement on the composite system. In quantum theory, this scenario removes the need for pre-shared entanglement between the parties.  We consider the framework of classical causal models \cite{Lee2019} to certify the entangled basis measurement of the receiver. In our tripartite PM scenario, two senders independently prepare quantum states according to their respective inputs and transmit them to a receiver, who performs a fixed joint measurement with $2^m$ outcomes on the received states.  We introduce suitable witnesses corresponding to the number of inputs and derive the bounds in the classical causal model. We show that the optimal quantum value of the witness exceeds the classical causal bound and uniquely certify the entangled basis measurement and the states sent by the senders.

The structure of the paper is as follows. In Sec.~\ref{sec:2}, we describe the PM scenario and the structure of the underlying classical causal model. In  Sec.~\ref{sec:3}, we propose SDI self-testing protocols for four-outcome maximally entangled measurements for the cases of binary inputs per sender and four inputs per sender, respectively. In Sec.~\ref{sec:5}, we provide a self-testing protocol for a 16-outcome entangled basis measurement whose projectors are maximally entangled across every $1|3$ bipartition.  In Sec.~\ref{sec:4}, we also examine the odd case $m=3$, where we establish a weaker self-testing result for  eight outcome entangled basis measurement  that ensures maximal entanglement across at least one bipartition. Next, in Sec.~\ref{sec:6}, we provide a generalized witness, the optimal quantum violation of which self-tests a $2^m$-outcome entangled basis measurement whose projectors are maximally entangled across every $1|(m-1)$ bipartition for even $m$.

\section{The PM scenario with two senders and one receiver \label{sec:2}}
A standard bipartite PM scenario involves two parties, one sender (Alice) and one receiver (Bob). Alice encodes the information she wishes to communicate into a quantum or classical system and sends it to Bob. Bob then attempts to recover the original information from the received message. When Bob has only one input, a $d$-dimensional classical system performs as well as a $d$-dimensional quantum system, as proved by Holevo \cite{Holevo1973} and subsequently generalized by Frenkel--Weiner \cite{Frenkel_2015}. Thus, no advantage can be obtained by using quantum systems. However, when Bob has more than one input, incompatible measurements on Bob's side can lead to a quantum advantage \cite{ambainis1998densequantumcodinglower,ambainis2009quantumrandomaccesscodes}. A natural question raises whether a quantum advantage can also be obtained by allowing multiple senders while restricting the receiver to a fixed input.
\begin{figure}[htbp]
    \centering
    \includegraphics[width=0.4\textwidth]{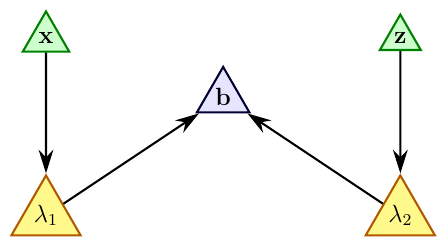}
    \caption{The DAG shown here represents a scenario involving two senders, Alice and Charlie, and one receiver, Bob. The inputs of Alice and Charlie are denoted by $\mathbf{x}$ and $\mathbf{z}$.  Based on their inputs, they prepare the states $\lambda_1$ and $\lambda_2$, respectively. Bob produces the outcome $\mathbf{b}$ depending on $\lambda_1$ and $\lambda_2$.}
    \label{fig:fig_1}
\end{figure}

In this paper,
we consider a PM scenario involving two senders(Alice and Charlie) and a single receiver(Bob). Each sender randomly receives arbitrary input and  based on their inputs, the senders prepare the corresponding states and transmit them to Bob. Bob then performs a fix joint measurement on the states received from the senders.  
A joint measurement is said to be non-classical if the correlations generated in the experiment are incompatible with the predictions of classical causal models. This incompatibility can be captured by an inequality in the probability space of the observed statistics, $f(\vec{P}) \leq \beta$, where  $\beta$ is a real number and  $\vec{P}$ denotes a probability vector whose dimension depends on the number of inputs and outputs of each party. In the following section, we show—using a non-linear witness inspired by \cite{Ghorai_et_al.PhysRevA.98.032110,sneha_et_al._2023_PhysRevA.108.062607} that, for the given scenario, there exists a joint measurement for bob  whose statistics can never be reproduced by any underlying classical causal model. The optimal quantum bound for the witness SDI self-tests the entangled  basis measurement of the receiver and the prepared states of the senders.

\begin{figure}[htbp]
    \centering
\begin{tikzpicture}[
    >=Stealth,
    node distance=1cm and 2cm,
    box/.style={
        draw,
        rectangle,
        minimum width=1.5cm,
        minimum height=1cm,
        align=center
    },
    AliceC/.style={
        box,
        fill=blue!12
    },
    BobC/.style={
        box,
        fill=orange!20
    },
    CharlieC/.style={
        box,
        fill=green!15
    },
    arrow/.style={->,thick},
    every node/.style={font=\normalsize}
]


\node[AliceC] (AliceC) {
    \textbf{Alice:}\\[1pt]
    Class. causal \\state\\[4pt]
    $\displaystyle P(\lambda_1\mid\mathbf{x})$
};

\node[BobC, right=.75cm of AliceC] (BobC) {
    \textbf{Bob:}\\[1pt]
    Class. resp.\\ function\\[1pt]
    $\displaystyle P(\mathbf{b}\mid\lambda_1,\lambda_2)$
};

\node[CharlieC, right=.75cm of BobC] (CharlieC) {
    \textbf{Charlie:}\\[1pt]
    Class. causal \\state\\[4pt]
    $\displaystyle P(\lambda_2\mid\mathbf{z})$
};

\node[anchor=east] at ([xshift=-0.2cm]AliceC.west)
    {\textbf{(a)}};

\draw[arrow] ($(AliceC.north)+(0,0.8)$)
    node[above] {$\mathbf{x}\equiv x_1x_2$}
    -- (AliceC.north);

\draw[arrow] ($(CharlieC.north)+(0,0.8)$)
    node[above] {$\mathbf{z}\equiv z_1z_2$}
    -- (CharlieC.north);

\draw[arrow] (AliceC.east) --
    node[above] {$\lambda_1$}
    (BobC.west);

\draw[arrow] (CharlieC.west) --
    node[above] {$\lambda_2$}
    (BobC.east);

\draw[arrow] (BobC.south) -- ++(0,-0.8)
    node[below] {$b\equiv b_1b_2\ldots b_m$};


\node[AliceC, below=3cm of AliceC] (AliceQ) {
    \textbf{Alice:}\\[4pt]
    Quan. state\\[6pt]
    $\displaystyle
    \rho^a_{\mathbf{x}}
    \in\mathcal{L}(\mathcal{H}_a)
    $
};

\node[BobC, right=0.75cm of AliceQ] (BobQ) {
    \textbf{Bob:}\\Joint quan. \\measurement\\[4pt]
    $\displaystyle \{E_b\}$
};

\node[CharlieC, right=0.75cm of BobQ] (CharlieQ) {
    \textbf{Charlie:}\\[4pt]
    Quan. state\\[6pt]
    $\displaystyle
    \rho^c_{\mathbf{z}}
    \in\mathcal{L}(\mathcal{H}_c)
    $
};

\node[anchor=east] at ([xshift=-0.2cm]AliceQ.west)
    {\textbf{(b)}};

\draw[arrow] ($(AliceQ.north)+(0,0.8)$)
    node[above] {$\mathbf{x}\equiv x_1x_2$}
    -- (AliceQ.north);

\draw[arrow] ($(CharlieQ.north)+(0,0.8)$)
    node[above] {$\mathbf{z}\equiv z_1z_2$}
    -- (CharlieQ.north);

\draw[arrow] (AliceQ.east) --
    node[above] {$\rho^a_{\mathbf{x}}$}
    (BobQ.west);

\draw[arrow] (CharlieQ.west) --
    node[above] {$\rho^c_{\mathbf{z}}$}
    (BobQ.east);

\draw[arrow] (BobQ.south) -- ++(0,-0.8)
    node[below] {$b \equiv b_1b_2\ldots b_m$};

\end{tikzpicture}
        \caption{
        (a) Represents a classical causal model of the scenario. Alice and Charlie prepare classical states, represented by $\lambda_1$ and $\lambda_2$, according to the conditional probability distributions $P(\lambda_1\mid\mathbf{x})$ and $P(\lambda_2\mid\mathbf{z})$, respectively. Bob produces the output $\mathbf{b}\equiv b_1b_2\ldots b_{m}$ according to the response function $P(\mathbf{b}\mid\lambda_1,\lambda_2)$, where $b_i\in\{0,1\}$ for all $i\in\{1,2\ldots,m\}$.
        (b) Quantum tripartite PM scenario, in which Alice and Charlie send the quantum states $\rho^a_{\mathbf{x}}$ and $\rho^c_{\mathbf{z}}$, respectively, to Bob. Bob then performs a joint quantum measurement, described by the positive operator-valued measure (POVM) $\{E_{\mathbf{b}}\}$, on the composite system received from the two senders.}\label{fig:classical-quantum}
\end{figure}
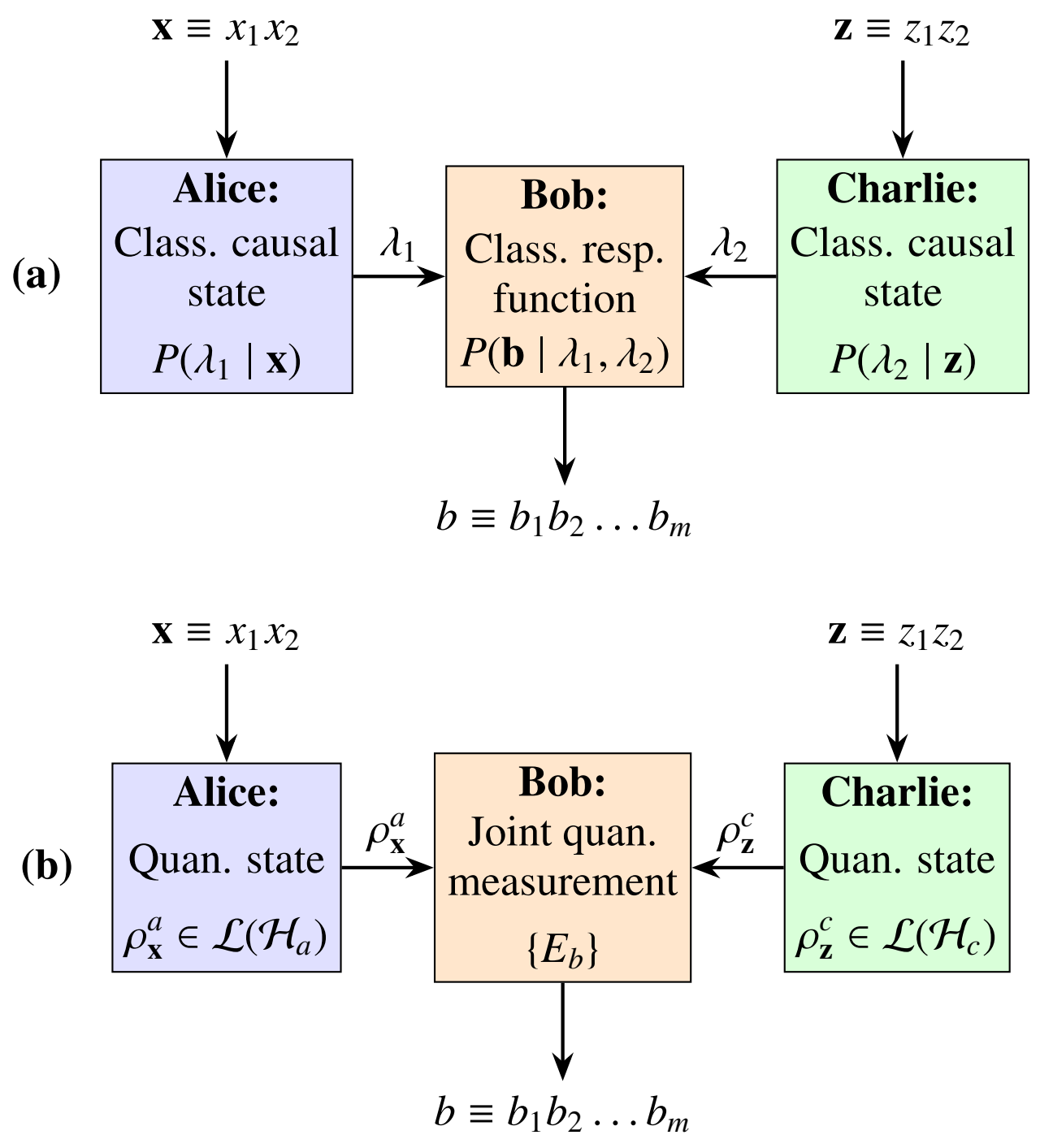

 
As depicted in Fig.~\ref{fig:classical-quantum}, we consider both senders to receive $2^{m}$ inputs where $m=2,3,4..$. We denote the inputs received by Alice and Charlie as \(\mathbf{x} \equiv x_1x_2\) and \(\mathbf{z} \equiv z_1z_2\), respectively, where \(x_1,z_1 \in \{0,1,.,2^{m-1}-1\}\) and \(x_2,z_2 \in \{0,1\}\). Based on their respective inputs, the senders prepare the corresponding states.
In the classical causal model, the state of the system is represented by a distribution \(P(\lambda)\) over the classical state \(\lambda\). In this scenario, the prepared state depends on the choice of input. Therefore, the distribution of the prepared state is conditioned on the input. We denote the probability distributions of the prepared states for Alice and Charlie by $P(\lambda_1|\mathbf{x})$ and $P(\lambda_2|\mathbf{z})$, respectively.
Upon receiving the states from both senders, Bob performs a $2^m$-outcome joint measurement, denoted by $\mathbf{b} \equiv b_1b_2\ldots b_{m}$, where $b_i \in \{0,1\} \,\forall\,\, i \in \{0,1,\cdots,m\}$.

Classical causal models are described by directed acyclic graphs (DAGs)~\cite{causal_assumptions_in_bell_theorem_PhysRevLett.114.140403,E._G._Cavalcanti_PhysRevX.8.021018,Wood_2015}. DAGs are widely used in Bayesian networks to represent the conditional dependencies among variables. In a DAG, variables are represented by nodes, while directed edges between the nodes indicate causal relationships. The acyclic property of the graph ensures that no vertex can be connected to itself through a directed path. The inputs $(\mathbf{x},\mathbf{z})$ of the preparing parties, the states of the physical systems $(\lambda_1,\lambda_2)$, and the measurement outcome $\mathbf{b}$ of the receiver are all treated as variables. The conditional dependencies among these variables are represented by the DAG shown in Fig.~\ref{fig:fig_1}, which corresponds to the scenario involving two senders, Alice and Charlie, and a single receiver, Bob.
 In Fig.~\ref{fig:fig_1}, $\lambda_1$ and $\lambda_2$ denote the states prepared on the basis of the inputs $\mathbf{x}$ and $\mathbf{z}$ provided to Alice and Charlie, respectively.

We assume that the sender parties prepare their states independently of one another; that is, the variables $\lambda_1$ and $\lambda_2$ are mutually independent. Consequently, the classical causal probability distribution of the composite state, on which the receiver performs a joint measurement,  depends only on the preparation inputs and can be written as
\begin{equation}
P(\lambda_1,\lambda_2 \mid \mathbf{x},\mathbf{z})
=
P(\lambda_1 \mid \mathbf{x})\, P(\lambda_2 \mid \mathbf{z}). \label{eq:s1}
\end{equation} 
In the classical causal model, the Bob's outcome probability can be decomposed as 
\begin{align}
P(\mathbf{b}|\mathbf{x},\mathbf{z})=&\iint d\lambda_1 d\lambda_2 P(\lambda_1|\mathbf{x})P(\lambda_2|\mathbf{z})P(\mathbf{b}|\lambda_1,\lambda_2)\label{eq:s2}
\end{align}
We demonstrate that the quantum statistics generated by certain entangled-basis measurements are incompatible with $P(\mathbf{b}\mid\mathbf{x},\mathbf{z})$ in Eq.~\eqref{eq:s2}. This certifies the nonclassicality of such entangled-basis measurements.

\section{self-testing of 
Four outcome Joint Measurement in case of four inputs per sender \label{sec:3} }
We consider that Alice and Charlie each randomly receive one of four possible inputs, represented by the two-bit strings $\mathbf{x}\equiv x_1x_2$ and $\mathbf{z}\equiv z_1z_2$, respectively, where $x_1,x_2,z_1,z_2\in\{0,1\}$, as depicted in Fig.~\ref{fig:classical-quantum}. Based on their inputs, they prepare states and send them to Bob. Bob then performs a single  joint measurement on the received states and produces four outcomes denoted by \( b_1 b_2 \in \{00, 01, 10, 11\} \). It gives rise to a joint probability distribution 
$
P(b_1b_2|\mathbf{x},\mathbf{z})$. 


We define a witness $\Delta^4$, with the corresponding classical causal bound and quantum bound denoted by $\Delta^4_c$ and $\Delta^4_Q$, respectively. 
\begin{align}
\Delta^4=\sum_{i=1}^{2} \sqrt{\left|K_{i}^{4}\right|} 
\label{eq:s3}
\end{align}
where the superscript $4$ in $K_i^4$ denotes the number of inputs received by the sender parties to prepare the state. 
We define $K_i^4$ as  
\begin{align}
K_{i}^{4}=\sum_{x_1,z_1=0}^{1} 
(-1)^{y_i^{x_1} +y_i^{z_1}}\sum_{x_2,z_2,b_1b_2}(-1)^{x_2+z_2+b_i}P(b_1b_2|\mathbf{x},\mathbf{z})
\label{eq:s4}
\end{align}
where, we define $y^{x_1} \equiv y_{i=1}^{x_1}y_{i=2}^{x_1}\in \{00, 01\}$ and   similarly for $y^{z_1} \equiv y_{i=1}^{z_1}y_{i=2}^{z_1}\in \{00, 01\}$. The $i^{th}$ bit of $y^{x_1}$ and $y^{z_1}$ is denoted by $y_i^{x_1},y_i^{z_1}$ respectively, i.e.,
$y^{x_{1}=1}_{i=1} = 0,\quad y^{x_{1}=1}_{i=2} = 0,\quad y^{x_{1}=2}_{i=1} = 0,\quad y^{x_{1}=2}_{i=2} = 1.$  The scheme we employ here to fix the value of $y_i^{x_1}$ and $y_i^{z_1}$  resembles the encoding scheme in random access codes \cite{Ghorai_et_al.PhysRevA.98.032110,ambainis2009quantumrandomaccesscodes}.
The distribution $P(b_1b_2|\mathbf{x},\mathbf{z} )$ generated by the the classical causal model satisfies Eq.\eqref{eq:s2}, Using this we derive the classical causal bound
\begin{align}
\Delta^4_C\leq 4
\label{eq:s5}
\end{align}
We provide a brief sketch of the proof here and a detailed proof is provided in Appendix \ref{app:appA1}.
 Using the probability distribution $P(b_1b_2|\mathbf{x},\mathbf{z})$ in Eq.\eqref{eq:s4}, we can write  
\begin{align}
K_{i}^{4}
&=
\iint d\lambda_1 d\lambda_2
\sum_{x_1,z_1}
(-1)^{y^{x_{1}}_i + y^{z_1}_i}
\sum_{x_2,z_2}
(-1)^{x_2+z_2}
P(\lambda_1|\mathbf{x})\notag \\
&\quad 
P(\lambda_2|\mathbf{z}) 
\sum_{b_1,b_2} (-1)^{b_i} P(b_1 b_2|\lambda_1,\lambda_2).
\label{eq:s5}
\end{align}
Since 
$
|\sum_{b_1,b_2} (-1)^{b_i} P(b_1 b_2|\lambda_1,\lambda_2)|\leq 1
$, we can further write  

\begin{align}
\Delta^4_C
\le
\sum_{i=1}^{2}& \sqrt{\int d\lambda_1 
\left|
\sum_{x_1x_2}
(-1)^{y^{x_1}_i+x_2}
P(\lambda_1|\mathbf{x})
\right|} \notag \\
& 
 \sqrt{\int d\lambda_2 \left|\sum_{z_1z_2}
(-1)^{y^{z_1}_i+z_2}
P(\lambda_2|\mathbf{z})
\right|} \label{eq:s6}
\end{align}
We prove that the above expression is bounded  $\Delta^{4}_{C} \leq 4$ in Appendix.~\ref{app:appA1}. 
 
In quantum theory, Alice and Charlie, each prepares four qubit states and send them to Bob. We denote the four qubit states prepared by Alice and Charlie as $\rho^a_{\mathbf{x}}\equiv\rho^a_{x_1 x_2}\in\{\rho^a_{00}, \rho^a_{01}, \rho^a_{10}, \rho^a_{11}\}$ and $\rho^{c}_{z}\equiv\rho^{c}_{z_1 z_2}\in\{\rho^{c}_{00}, \rho^{c}_{01}, \rho^{c}_{10}, \rho^{c}_{11}\}$, respectively. 
Bob performs a four-outcome joint measurement described by a POVM denoted by  $E_{b_1 b_2}\in \{E_{00}, E_{01}, E_{10}, E_{11}\}$, where $E_{b_1 b_2}\geq 0$ and $\sum_{b_1b_2}E_{b_1b_2}=\mathbb{I}$.
 Since both senders prepare their states independently, we can write the composite state Bob receives as $\rho_{\mathbf{x},\mathbf{z}}=\rho^a_{\mathbf{x}} \otimes \rho^c_{\mathbf{x}}$. The quantum probability distribution can be derived as 
\begin{eqnarray}
    P_Q(b_1b_2|\mathbf{x},\mathbf{z}) =\text{Tr}\left[(\rho^a_{\mathbf{x}} \otimes \rho^c_{\mathbf{z}}) E_{b_1b_2}\right]\label{eq:s7}  
\end{eqnarray}
Now, by inserting Eq. (\ref{eq:s7}) in Eq.\eqref{eq:s4} and by simplifying we get  
\begin{align}
    K_{1}^{4} = \text{Tr}[&(\rho^a_{00} - \rho^a_{01} + \rho^a_{10} - \rho^a_{11}) \notag\\\otimes &(\rho^c_{00} - \rho^c_{01} + \rho^c_{10} - \rho^c_{11}) 
     B_1]\notag\\
K_2^{4} = \text{Tr}[&(\rho^a_{00} - \rho^a_{01} - \rho^a_{10} + \rho^a_{11}) \notag \\ \otimes &(\rho^c_{00} - \rho^c_{01} - \rho^c_{10} + \rho^c_{11}) B_2]\label{eq:s8}
\end{align}
where  $B_1$ and $ B_2$ are 
\begin{align}
    B_1=E_{00}-E_{01}-E_{10}+E_{11}\notag
 \\B_2= E_{00} - E_{01} + E_{10} -E_{11}\label{eq:s9}
 \end{align}
Further, we consider \begin{align}
\frac{\rho^a_{00} - \rho^a_{01} + \rho^a_{10} - \rho^a_{11} }{\omega^a_1}=A_1\, ,\,\, \frac{\rho^c_{00} - \rho^c_{01} + \rho^c_{10} - \rho^c_{11}}{\omega^c_1}= C_{1}\nonumber \\ 
\frac{\rho^a_{00} - \rho^a_{01} - \rho^a_{10} + \rho^a_{11}}{\omega^a_2} =A_2\, ,\,\, \frac{\rho^c_{00} - \rho^c_{01} - \rho^c_{10} + \rho^c_{11}}{\omega^c_2}= C_{2}  \label{eq:s10}
\end{align}
where $\omega^a_i$ is the normalized Hilbert-Schmidt norm. For an operator O, the normalized Hilbert-Schmidt norm is defined as follows 
\begin{eqnarray}
    \omega = \frac{\sqrt{Tr(O^\dagger O})}{\sqrt{d}} \label{eq:s10.5}
\end{eqnarray}
where $d$ is the dimension of the system.
Now, by using Eqs.\eqref{eq:s9} and \eqref{eq:s10}, the quantum value of $\Delta^{4}$ can be re-written as
\begin{eqnarray}
\Delta^{4}_{Q} =   \sqrt{|\omega^a_{1} \omega_{1}^{c}\text{Tr}[(A_1\otimes C_1)  B_1]|}+ \sqrt{|\omega^a_{2} \omega_{2}^{c}\text{Tr}[(A_2\otimes C_2)  B_2]|}
\end{eqnarray}
which can be further written as  
\begin{align}
\Delta^{4}_{Q}  \leq 2 \bigg(\sqrt{|\omega^a_1  \omega^c_1|}+\sqrt{|\omega^a_2 \omega^c_2|}\bigg) \label{eq:s12}
\end{align}
Note that equality holds in Eq.\eqref{eq:s12},when
\begin{eqnarray}
Tr[(A_1\otimes C_1)  B_1]=Tr[(A_2\otimes C_2)  B_2]=4
\end{eqnarray}
Using the inequality $\sqrt{\omega^a_1\omega^c_1}+\sqrt{\omega^a_2\omega^c_2} \le \sqrt{\omega^a_1 + \omega^a_2}\,\sqrt{\omega^c_1 + \omega^c_2}$ together with the inequality $
\omega^a_1+\omega^a_2 \leq \sqrt{2((\omega^a_1)^{2}+(\omega^a_2)^2)}$
, we can write the quantum upper bound as
\begin{align}
      \Delta^{4}_{Q} \leq  2\bigg(\sqrt[4]{2  {((\omega^a_1)^2+(\omega^a_2)^2)}} \cdot \sqrt[4]{2  \big((\omega^c_1)^2+(\omega^c_2)^2\big)} \bigg)\label{eq:s13}
\end{align}
Calculating the norm of each operator and adding the squares of them yields
\begin{align}
(\omega^a_1)^2+(\omega^a_2)^2=\text{Tr}\big[(\rho^a_{00})^2+(\rho^a_{01})^2+(\rho^a_{10})^2+(\rho^a_{11})^2\notag\\-2(\{\rho^a_{00},\rho^a_{01}\}+\{\rho^a_{10},\rho^a_{11}\})\big]\label{eq:s15}
\end{align}
and a similar expression appears for $(\omega^c_1)^2+(\omega^c_2)^2$. This 
 then yields 
\begin{align}
 \Delta^{4}_{Q}\leq 4\sqrt{2} \label{eq:s16}
\end{align}
with maximization conditions  
\begin{align}
&\text{Tr}[(\rho^a_{\mathbf{x}})^2]=\text{Tr}[(\rho^c_{\mathbf{z}})^2]=1 \,\forall x, z\in\{00,01,10,11\} \notag\\ 
& \text{Tr}[\rho^a_{00}\rho^a_{01}]=\text{Tr}[\rho^a_{10}\rho^a_{11}]=\text{Tr}[\rho^c_{00}\rho^c_{01}]=\text{Tr}[\rho^c_{10}\rho^c_{11}]=0.\label{eq:s17}
\end{align}
Therefore, the maximization restricts the prepared states of Alice and Charlie to be pure qubit states satisfying the pairwise orthogonality conditions $\rho^a_{00} \perp \rho^a_{01}$ and  $\rho^a_{10} \perp \rho^a_{11}$, as well as $\rho^{c}_{00} \perp \rho^{c}_{01}$ and $\rho^{c}_{10} \perp \rho^{c}_{11}$.
Furthermore, the right-hand sides of Eqs.\eqref{eq:s12} and \eqref{eq:s13} become equal when $\omega^a_1=\omega^a_2$  and $\omega^c_1=\omega^c_2$. This implies
\begin{align}
\text{Tr}(\rho^a_{00} \rho^a_{10}) = 
\text{Tr}(\rho^a_{00} \rho^a_{11}) &= \frac{1}{2}\notag\\
\text{Tr}(\rho^a_{01} \rho^a_{10}) =
\text{Tr}(\rho^a_{01} \rho^a_{11}) &= \frac{1}{2} \label{eq:s18}
\end{align}
and consequently, $\omega^a_1=\omega^a_2=\omega^c_1=\omega^c_2=\sqrt{2}$. Maximum quantum bound in \eqref{eq:s12} is achived when $\text{Tr}[(A_i \otimes C_i)(B_i)] =4$ for all $i=1,2.$ Using Eqs.\eqref{eq:s17} and \eqref{eq:s18}, we obtain  $(A_1)^2=(A_2)^2=(C_1)^2=(C_2)^2=I_2$ along with $\{A_1,A_2\}=\{C_1,C_2\}=0$. For a tight quantum bound this implies 
\begin{align}
 A_i \otimes C_i=B_i \,,\,\,
(B_i)^2=I_2 \otimes I_2  \, , 
[B_1,B_2]=0 \label{eq:s20}
\end{align}
Altogether, the equality holds in Eq.~\eqref{eq:s12}, so that $(\Delta^{4}_{Q})^{opt}= 4\sqrt{2} $.

We now prove that $(\Delta^{4}_{Q})^{opt}$ certifies that Bob's measurement operators $\{E_{b_1b_2}\}$ are maximally entangled projectors. We start By squaring both sides of Eq.~\eqref{eq:s9} and, using Eq.~\eqref{eq:s20} together with the completeness condition of the POVMs, we obtain
\begin{subequations}\label{eq:s21}
\begin{eqnarray}
    (E_{00}+E_{11})(E_{01}+E_{10})=0\label{eq:first}\\
    (E_{00}+E_{10})(E_{01}+E_{11})=0\label{eq:second}\\
    (E_{00}-E_{01})(E_{10}-E_{11})=0\label{eq:third}
    \end{eqnarray}
\end{subequations}
Consider the operators $E_{01}+E_{10}=P,\,\,  E_{00}+E_{11}=Q.
$
Since $E_{01},E_{10}\geq 0$, it follows that
\begin{eqnarray}
0 \leq E_{01} \leq P. \label{eq:s23}
\end{eqnarray}
Applying the congruence transformation with $Q$ to Eq.~\eqref{eq:s23}, we obtain
\begin{eqnarray}
0 \leq QE_{01}Q \leq QPQ. \label{eq:s24}
\end{eqnarray}
From Eq.~\eqref{eq:first}, we have $PQ=0$. Hence,
\begin{eqnarray}
QE_{01}Q=0. \label{eq:s25}
\end{eqnarray}
Since $E_{01}$ is positive semidefinite, it admits the decomposition
$
E_{01}=E_{01}^{1/2}E_{01}^{1/2}.
$
Substituting this into Eq.~\eqref{eq:s25}, we obtain
 \begin{eqnarray}
    ( E_{01}^{\frac{1}{2}}Q)^\dagger (E_{01}^{\frac{1}{2}}Q)=0\label{eq:s26}.
 \end{eqnarray}
This implies $E_{01}Q=0$. since $P+Q=I$,
we can write $[E_{01},P]=0$ which gives, $[E_{01},E_{10}]=0$.
The same mathematical steps can be applied to Eq.~\eqref{eq:second}. Thus, we obtain
\begin{eqnarray}
    [E_{00},E_{11}]=[E_{00},E_{10}]=[E_{01},E_{11}]=[E_{01},E_{10}]=0.
    \label{eq:s28}
\end{eqnarray}
Using the completeness relation of the POVM together with Eq.~\eqref{eq:s28}, we further obtain
$[E_{00},E_{01}]=[E_{10},E_{11}]=0$.
The product of two commuting positive semidefinite operators is also positive semidefinite. Therefore, from Eqs.~\eqref{eq:first} and \eqref{eq:second}, we obtain
\begin{eqnarray}
E_{00}E_{01}=E_{00}E_{10}=E_{11}E_{01}=E_{11}E_{10}=0.
\end{eqnarray}
Combining these relations with Eq.~\eqref{eq:third}, we obtain the remaining two orthogonality relations,
$E_{01}E_{10}=E_{00}E_{11}=0$.
The idempotency of the POVM elements follows directly from the orthogonality relations together with the completeness relation.  
Hence, the POVM elements $\{E_{b_1b_2}\}$ are projectors $\{\pi_{b_1 b_2}\}$ satisfying both orthogonality and idempotency.

We next show that achieving the maximum quantum bound self-tests the measurement projectors $\{\pi_{b_1 b_2}\}$ to be maximally entangled.
From the definition of $B_1 \, \text{and} \, B_2$ given in Eq.\eqref{eq:s9},  measurement projectors can then be written as 
\begin{align}
\pi_{b_1 b_2} &= \tfrac{1}{4}\big( I + (-1)^{b_1}B_{1} + (-1)^{b_2}B_{2}+(-1)^{b_1+b_2}B_{3} \big)
\end{align}
By taking partial trace of first subsystem, we have
\begin{eqnarray}
\mathrm{Tr}_1[\pi_{b_1 b_2}] &=& \frac{1}{4} \Big[ 
\mathrm{Tr}_1[\mathbb{I}_2 \otimes \mathbb{I}_2] 
+ (-1)^{b_1}\mathrm{Tr}_1[B_1]\\
\nonumber
&&+(-1)^{b_2} \mathrm{Tr}_1[B_2] 
+ (-1)^{b_1 b_2}\mathrm{Tr}_1B_3] 
\Big]
\end{eqnarray}
From Eq.\eqref{eq:s10}, it is easy to find that
$
\mathrm{Tr}_1[B_2] = \mathrm{Tr}_1[B_1] = 0$. Also, $Tr_{1}[B_3]=Tr_{1}[B_1 B_2]=0$. Therefore, $\mathrm{Tr}_1[\pi_{b_1 b_2}]=\mathbb{I}_2/2$\ \  $\forall \,b_1,b_2 \in\{0,1\}$. Similarly, it can be shown that $\mathrm{Tr}_2[\pi_{b_1 b_2}]=\mathbb{I}_2/2$,\ $\forall \,b_1,b_2 \in\{0,1\}$. Therefore, every measurement projector has maximally mixed single-qubit marginals and consequently, the projectors $\{\pi_{b_1 b_2}\}$ are maximally entangled. Hence, the optimal quantum bound $\Delta^4_Q=4\sqrt{2}$,  self-test Alice and Charlie's states and the joint measurement projectors of Bob to be maximally entangled.

An example of a set of prepared states and an entangled measurement that attain the optimal quantum value is given below. 
\begin{align}\
&\rho_{x_1x_2}=\frac{1}{2}(\mathbb{I}_2+(-1)^{x_2}\frac{\sigma_x+(-1)^{x
   _1}\sigma_z}{\sqrt{2}}) \,\, \forall x_1,x_2 \in\{0,1\} \notag\\ 
   &\rho_{z_1z_2}=\frac{1}{2}(\mathbb{I}_2+(-1)^{z_2}\frac{\sigma_x+(-1)^{z
   _1}\sigma_z}{\sqrt{2}}) \,\, \forall z_1,z_2 \in\{0,1\}\notag\\
   &\{\pi_{b_1b_2}\}=\left\{
|\phi^+\rangle\langle\phi^+|,
|\phi^-\rangle\langle\phi^-|,
|\psi^+\rangle\langle\psi^+|,
|\psi^-\rangle\langle\psi^-|
\right\}
\end{align}
Here, $\{|\phi^+\rangle\langle\phi^+|,
|\phi^-\rangle\langle\phi^-|,
|\psi^+\rangle\langle\psi^+|,
|\psi^-\rangle\langle\psi^-|\}$ is bell basis.
\section{Self-testing of 
Four outcome measurements for the case of eight inputs per sender \label{sec:4}}
In this scenario, Alice and Charlie receives eight possible inputs denoted by  $\mathbf{x} \equiv x_1x_2$ and $ \mathbf{z} \equiv z_1z_2$  where $x_1,z_1 \in \{0,1,2,3\}$ and $x_2,z_2 \in \{0,1\}$ respectively. Bob performs a four-outcome joint measurement. Similar to the previous case, we assume preparation independence between Alice and Charlie, the joint probability distribution $P(b_1b_2|\mathbf{x},\mathbf{z})$ of the underlying causal model is decomposed as Eq.\eqref{eq:s2}.
To self-test the entangled measurement, we consider the following witness $\Delta^8$ given as  
\begin{align}
\ \Delta^8=\sum_{i=1}^{3} \sqrt{\left|K_{i}^8\right|} 
\label{eq:s31}
\end{align}
where $K_i^8$ is defined as follows
\begin{align} K^8_{i}=\sum_{x_1,z_1=0}^{3}
(-1)^{y_i^{x_1} +y_i^{z_1}}\sum_{x_2,z_2,b_1b_2}(-1)^{x_2+z_2+b_i}P(b_1b_2|x_1x_2,z_1z_2) \label{eq:s32}
\end{align}
where, $y_i^{x_1},y^{z_1}_i$ both  can take value 0 or 1. To fix the value of $y^{x_1}_i,y^{z_1}_i$, we define $y^{x_1}$ and $y^{z_1}$ as a 3-bit sting belongs to set $\{000,001,010,100\}$. The 
$i^{th}$ bit of $y^{x_1}$ and $y^{z_1}$ is denoted by $y_i^{x_1}$ and $y_i^{z_1}$, respectively. For example if $x_1=1$ we have $y^{x_1=1}=001,$ so $y^{1}_{i=1}=0 ,y^{1}_{i=2}=0  ,y^{1}_{i=3}=1.$
The expression in  Eq.\eqref{eq:s31} is similar to 
Eq.\eqref{eq:s3}; the only difference lies in the number of inputs. Specifically, now the summations is over $x_1$ and $z_1$  range from $0$ to $3$, while the summation over $i$ runs from $1$ to $3$.
Therefore, in the calculation of the classical bound, only the terms $\sum_{x_1,z_1}$ and $\sum_i$ differ from the previous case. Hence, using Eq.~\eqref{eq:s6}, we can write: 
\begin{align}
\Delta^8_C
\le
\sum_{i=1}^{3}& \sqrt{\int d\lambda_1 
\left|
\sum_{x_1 =0}^{3}
(-1)^{y_i^{x_1}} \sum_{x_2=0}^{1} (-1)^{x_3}
P(\lambda_1|x_1x_2)
\right|} \notag \\
& 
 \sqrt{\int d\lambda_2 \left|
\sum_{z_1=0}^{3}
(-1)^{y_i^{z_1}}\sum_{z_2=0}^{1} (-1)^{z_2}
P(\lambda_2|z_1z_2)
\right|}\label{eq:s33}  
\end{align}
We prove in detail in  Appendix.~\ref{app:A2}, the above expression is bounded by $\Delta_C^8 \leq 12$. 
 
Next, we consider the quantum case, in which Alice and Charlie each prepare eight quantum states, denoted by $\rho^a_{\mathbf{x}}$ and $\rho^c_{\mathbf{z}}$, respectively where $\mathbf{x}\equiv x_1x_2 \, , \mathbf{z}=z_1z_2$. We assume the states prepared by both senders to be qubit states. Bob performs four-outcome joint POVMs measurement $\{E_{b_1b_2}\}$, where $b_1,b_2 \in \{0,1\}$, on the composite state $\rho_{\mathbf{x},\mathbf{z}}$ received from the senders. Under the preparation-independence assumption, the composite state can be written as $\rho_{\mathbf{x},\mathbf{z}} = \rho^a_{\mathbf{x}} \otimes \rho^c_{\mathbf{z}}$.
The joint quantum probability distribution can be written as $P( b_1b_2|\mathbf{x},\mathbf{z}) =\text{Tr}(\rho_{\mathbf{x},\mathbf{z}} E_{b_1b_2})$. Putting this into Eq.\eqref{eq:s32},  $K_i^8$ becomes
\begin{align}
K_i^8=Tr\left[\sum_{x_1,z_1 =0}^{3}(-1)^{y_i^{x_1}+y_i^{z_1}} \sum_{x_2,z_2=0}^{1}(-1)^{{x_2}+{z_2}} (\rho^a_{\mathbf{x}} \otimes \rho^{c}_{\mathbf{z}})B_i\right]\label{eq:s34}
\end{align}
 where $B_i$ is defined as follows 
 \begin{eqnarray}
     B_i=\sum_{b_1,b_2=0}^{1} (-1)^{b_i} E_{b_1b_2}
\,\,\text{where}\,\, b_{i=3}=b_1 \oplus b_2 \label{eq:s34.5}
 \end{eqnarray} 
For notational convenience, we define the operators $A_i$ and $C_i$ as
\begin{align}
A_i=\frac{1}{\omega^a_{i}}\sum_{x_1=0}^{3}(-1)^{y_i^{x_1}} \sum_{x_2=0}^{1}(-1)^{x_2} \rho^a_{x_1x_2}\notag \\ 
C_i=\frac{1}{\omega^c_i}\sum_{z_1=0}^{3}(-1)^{y_i^{z_1}} \sum_{z_2=0}^{1}(-1)^{z_2} \rho^{c}_{z_1z_2}\label{eq:s35}
\end{align}
where $\omega^a_i$ and $\omega^c_i$ denote the  normalized Hilbert-Schmidt norm defined in Eq.~\eqref{eq:s10.5}. Using Eq.~\eqref{eq:s35}, the quantum value can then be written as
\begin{align}
\Delta^8_Q =\sum_{i=1}^{3} \sqrt{ \left| \omega^{a}_i \omega^{c}_i Tr[(A_i \otimes C_i) B_i]\right| } \label{eq:s36}
\end{align}
Since $\max \operatorname{Tr}[(A_i \otimes C_i)B_i]=4$ for all $i=1,2,3$, we can further write
\begin{eqnarray}
    \Delta^8_Q \leq  2\bigg( \sqrt{\omega^a_1 \omega^c_1}+\sqrt{\omega^a_2 \omega^c_2}+\sqrt{\omega^a_3 \omega^c_3} \ \bigg) \label{eq:s37}
\end{eqnarray} 
Using  the inequality 
$ \sqrt{r_1s_1}+\sqrt{r_2s_2}+\sqrt{r_3s_3}
\leq \sqrt{r_1+r_2+r_3}\cdot\sqrt{s_1+s_2+s_3} 
$ together with the inequality $
r_1+r_2+r_3 \leq \sqrt{3((r_1)^{2}+(r_2)^2+(r_3)^2)}
$ for all $r_1, r_2,r_3, s_1, s_2,s_3 \geq 0$,the quantum bound can be expressed as 
\begin{align}
\Delta^8_Q \leq 2\sqrt[4]{9 \bigg((\omega^a_1)^2+(\omega^a_2)^2+(\omega^a_3)^2 \bigg)  \bigg((\omega^c_1)^2+(\omega^c_2)^2+(\omega^c_3)^2\bigg)} \label{eq:s38}
\end{align}
By using the definition of  normalized Hilbert-Schmidt norm in Eq.\eqref{eq:s10.5} we can write the following. 
\begin{align}
\sum_{i=1}^{3}(\omega^a_i)^2=\frac{1}{2}\text{Tr}\bigg(\sum_{x_1=0}^{3}\sum_{x_2=0}^{1} 4(\rho^a_{x_1x_2})^2-2 \sum_{x_1=0}^{3}\{\rho_{x_10},\rho_{x_11}\}-(\delta^a)^2 \bigg) \label{eq:s39}
\end{align}
where 
$\delta^a=(\rho^a_{00}-\rho^a_{01}-\rho^a_{10}+\rho^a_{11}-\rho^a_{20}+\rho^a_{21}-\rho^a_{30}+\rho^a_{31})
$.
A detailed calculation of the above step is provided in Appendix~\ref{app:appendixB_1}. 
The same calculation can be carried out for $\sum_{i=1}^3(\omega^c_i)^2$. 
To maximize the expression in Eq.~\eqref{eq:s39}, all terms with negative coefficients are set to zero, while the terms with positive coefficients attain their maximum value, i.e.,
\begin{align}
&Tr[(\rho^a_{x_1x_2})^2]=Tr[(\rho^c_{z_1z_2})^2]=1\notag\\
&Tr[\rho^a_{x_10}\rho^a_{x_11}] 
=Tr[\rho^c_{z_10}\rho^c_{z_11}]=0\notag\\
&\delta^a=\delta^c=0\, \forall\,x_1,z_1 \in (0,1,2,3) ,x_2,z_2 \in(0,1)
\label{eq:s40}
\end{align}
This implies
$\sum_{i=1}^{3}(\omega^a_i)^2=\sum_{i=1}^{3}(\omega^c_i)^2=16$.
Hence, we obtain the quantum upper bound
$\Delta^8_Q \leq 8 \sqrt{3}$.
The equality between the right hand side of Eq.~\eqref{eq:s37} and Eq.~\eqref{eq:s38} holds when $\omega^a_1=\omega^a_2=\omega^a_3$ and $\omega^c_1=\omega^c_2=\omega^c_3$.
Using this equality of norms, together with $\delta^a=\delta^c=0$, we derive the following relation between the non-orthogonal states
\begin{align}
&Tr[(\rho^a_{00}-\rho^a_{01})(\rho^a_{10}-\rho^a_{11})]=
Tr[(\rho^a_{00}-\rho^a_{01})(\rho^a_{20}-\rho^a_{21})]=\frac{2}{3},\notag\\
&Tr[(\rho^a_{00}-\rho^a_{01})(\rho^a_{30}-\rho^a_{31})]=\frac{2}{3},
Tr[(\rho^a_{10}-\rho^a_{11})(\rho^a_{20}-\rho^a_{21})]=\frac{-2}{3},\notag\\
&Tr[(\rho^a_{10}-\rho^a_{11})(\rho^a_{30}-\rho^a_{31})]=
Tr[(\rho^a_{20}-\rho^a_{21})(\rho^a_{30}-\rho^a_{31})]=\frac{-2}{3}\label{eq:s41}
\end{align}
Similar relations also hold for Charlie's prepared states. A detailed derivation of the above relations is provided in Appendix~\ref{app:appendixB_1}. Using the maximization conditions in Eqs.~\eqref{eq:s40} and \eqref{eq:s41}, we obtain $(A_i)^2=(C_i)^2=\mathbb{I}_2 \,\, \forall \,i\in\{1,2,3\}$. The equality in Eq.~\eqref{eq:s37} is achieved when $\operatorname{Tr}[(A_i \otimes C_i)B_i]=4$ for all i=1,2,3. Since we have already established that $A_i$ and $C_i$ are involutory operators, the quantum bound is tight when 
\begin{eqnarray}
    A_i \otimes C_i=B_i\,\, ,\,\, (B_i)^2=I_4 \label{eq:s42}
\end{eqnarray}
Next, we show that, to achieve the quantum maximum $(\Delta^8_{Q})^{opt}=8\sqrt{3}$, the measurement operators $\{E_{b_1b_2}\}$ must be a set of maximally entangled projectors $\{\pi_{b_1b_2}\}$. Using Eq.~\eqref{eq:s34.5}, we can write the measurement projector as 
\begin{align}
\pi_{b_1 b_2} &= \tfrac{1}{4}\big( \mathbb{I}_2 + (-1)^{b_1}B_{1} + (-1)^{b_2}B_{2}+(-1)^{b_1 \oplus b_2}B_{3} \big) \label{eq:s44}
\end{align}
It follows from Eq.~\eqref{eq:s35}, that \(A_i\) and \(C_i\) are traceless operators for all \(i\). This, together with the decomposition in Eq.~\eqref{eq:s42}, implies that the reduced operators obtained by tracing out either subsystem vanish:
$
\text{Tr}_1(B_i)=\text{Tr}_2(B_i)=0,
\, \forall\,i,
$
which gives  
\begin{eqnarray}
   Tr_1(\pi_{b_1b_2})=Tr_2(\pi_{b_1b_2})=\frac{\mathbb{I}_2}{2}
\end{eqnarray}
Therefore, $(\Delta_Q^8)^{opt}=8 \sqrt{3}$, self-tests the measurement to be maximally entangled projectors and the states prepared by Alice and Charlie to be pure qubit states satisfying the pairwise orthogonality and trace relations given in Eq.~\eqref{eq:s41}. An explicit example of a set of prepared state and entangled measurement that attain the optimal quantum value is provided in Appendix~\ref{app:appendixB_1}.  
\section{Self-testing of 
Eight outcome entangled measurement \label{sec:5}}
To self-test the eight-outcome entangled measurement, we employ the similar witness introduced in Sec.~\ref{sec:3} (see Eqs.~\eqref{eq:s31} and \eqref{eq:s32}), with two key modifications. First, Bob performs an eight-outcome measurement, denoted by ${b_1b_2b_3}$, where $b_i\in\{0,1\}$ for all $i=1,2,3$. Second, the dimensions of the systems prepared by Alice and Charlie are four and two, respectively. The classical causal bound in the present scenario is $
\Delta^8_C\leq 12.
$ See Appendix~\ref{app:A2} for details.

In the quantum case, the states prepared by Alice and Charlie are denoted by $\rho^a_{\mathbf{x}}\equiv\rho^a_{x_1x_2}\in\mathbb{C}^4$ and $\rho^c_{\mathbf{z}}\equiv\rho^c_{z_1z_2}\in\mathbb{C}^2$, respectively. Bob's eight-outcome projective measurement is denoted by $\{\pi_{b_1b_2b_3}\}$, where $b_i\in\{0,1\}$ with $i=1,2,3$. To derive the quantum bound, we define $A_i$ and $C_i$ analogously to Eq.~\eqref{eq:s35}. Following the similar steps as in Sec.~\ref{sec:3}, we obtain the quantum bound as
\begin{align}
\Delta^Q =\sum_{i=1}^{3} \sqrt{ \left| \omega^{a}_i \omega^{c}_i Tr[(A_i \otimes C_i) B_i]\right| } \label{eq:s46}
\end{align}
where $B_i=\sum_{b_1,b_2,b_3}(-1)^{b_i}\pi_{b_1b_2b_3}$. Since $\max\operatorname{Tr}[(A_i\otimes C_i)B_i]=8$ for all i=1,2,3, we can further write
\begin{eqnarray}
    \Delta^8_Q \leq  2\sqrt2\bigg( \sqrt{\omega^a_1 \omega^c_1}+\sqrt{\omega^a_2 \omega^c_2}+\sqrt{\omega^a_3 \omega^c_3} \ \bigg) \label{eq:s47}
\end{eqnarray} 
Following the same procedure as in Sec.~\ref{sec:3}, the quantum bound can be written as
\begin{align}
\Delta^Q \leq 2\sqrt2\sqrt[4]{9 \bigg((\omega^a_1)^2+(\omega^a_2)^2+(\omega^a_3)^2 \bigg)  \bigg((\omega^c_1)^2+(\omega^c_2)^2+(\omega^c_3)^2\bigg)} \label{eq:s38}
\end{align}
We derive the maximum values of the expressions $\sum_{i=1}^{3}(\omega_i^a)^2=4$ and $\sum_{i=1}^{3}(\omega_i^c)^2=16$. A detailed calculation is provided in Appendix~\ref{App:b3}. The difference between these two values arises from the different dimensions of the systems prepared by Alice and Charlie. Consequently, the optimal quantum bound is $(\Delta^8_Q)^{\mathrm{opt}}=8\sqrt{3}$.
The optimal quantum bound is achieved when $\omega_1^a=\omega_2^a=\omega_3^a=\frac{2}{\sqrt{3}}$ and $\omega_1^c=\omega_2^c=\omega_3^c=\frac{4}{\sqrt{3}}$. Furthermore, we derive 
\begin{align}
B_i=\frac{3}{8}(A_i\otimes C_i),\, \{A_i,A_j\} = \{C_i,C_j\}=0\,\, \forall i\neq j.\;\label{eq:s39}
\end{align}
Using the maximization condition given in Eq.~\eqref{eq:s39}, for each measurement projector $\{\pi_{b_1b_2b_3}\}$, we find that the reduced operator of the third subsystem, obtained by tracing out the first two subsystems, is maximally mixed. Therefore, for each outcome, the corresponding measurement projector $\pi_{b_1b_2b_3}$ represents a state that is maximally entangled across the \(12|3\) bipartition. See Appendix~\ref{App:b3} for a detailed derivation.
 \section{Self testing of 16 outcome joint measurement\label{sec:6}}
To self-test the sixteen outcome joint measurement, we consider two senders, Alice and Charlie, each receiving 16 inputs denoted by $\mathbf{x} \equiv x_1x_2$ and $\mathbf{z} \equiv z_1z_2$ respectively, where $x_1$ and $z_1$ $\in \{0,1,2,3,4,5,6,7\}$ and $x_2$,$z_2 \in \{0,1\}$. Based on their inputs, they prepare states and send them to Bob. Bob then performs a fixed sixteen-outcome joint measurement denoted by $\mathbf{b}=b_1b_2b_3b_4$, where $b_i \in \{0,1\}$  on the composite system received from Alice and Charlie. 
In the case of the classical causal model, the joint probability distribution  is given by
\begin{align}
P(b_1b_2b_3b_4|\mathbf{x},\mathbf{z})=\int d\lambda_1  d\lambda_2  P(b_1b_2b_3b_4| \lambda_1 ,\lambda_2)P(\lambda_1|\mathbf{x}) P(\lambda_2|\mathbf{z}) \label{eq:s46}
\end{align}
We employ the following  witness to self-test the 16-outcome entangled measurement 
\begin{align}
\Delta^{16}=\sum_{i=1}^{4} \sqrt{\left|K_{i}^{16}\right|} \label{eq:s47}
\end{align}
where $K_i^{16}$ is defined as 
\begin{align} 
K_{i}^{16}=&\sum_{x_1,z_1=0}^{7}
(-1)^{y_i^{x_1} +y_i^{z_1}}\sum_{x_2z_2b_1b_2b_3b_4}(-1)^{x_2+z_2} (-1)^{b_i}\notag \\&P(b_1b_2b_3b_4|\mathbf{x},\mathbf{z}) \label{eq:s49}
\end{align}
Here, $y^{x_1},y^{z_1}\in \{0000,0001, 0010, 0100,1000,0011,0101,\\0110\}$. Each  $y^{x_1},y^{z_1}$ is a 4-bit string. Similar to Sec.\ref{sec:2} and \ref{sec:3} the $i^{th}$ bit of $y^{x_1}$ and $y^{z_1}$ is denoted by $y_i^{x_1},y_i^{z_1}$ respectively.

The probability distribution $P(b_1b_2b_3b_4|\mathbf{x},\mathbf{z})$ generated by a classical causal bound gives
\begin{align}
\Delta^{16}_C \leq 24 \label{eq:s47}
\end{align}
A detailed derivation of the classical causal model bound is provided in Appendix~\ref{app:A2}.

Next, we consider the quantum case, in which Alice and Charlie's preparation are denoted by $\rho^a_\mathbf{x} \equiv \rho^a_{x_1x_2}$ and $\rho^c_\mathbf{z}=\rho^c_{z_1z_2}$ respectively, where $x_1,z_1 \in \{0,1,2,3,4,5,6,7\}$ and $x_2,z_2 \in \{0,1\}.$ We assume that the dimension of both prepared systems is four. Bob then performs a sixteen-outcome projective measurement denoted by $\{\pi_{b_1b_2b_3b_4}\}$ where $\sum_{b_1,b_2,b_3,b_4=0}^{1}\pi_{b_1b_2b_3b_4}=\mathbb{I}_{16}$. We can write the joint probability distribution as  
$P(b_1b_2b_3b_4|\mathbf{x},\mathbf{z})=\text{Tr}(\rho^a_{
\mathbf{x}} \otimes \rho^c_{\mathbf{z}} \pi_{b_1b_2b_3b_4}) $. 
By putting the value of $P(b_1b_2b_3b_4|\mathbf{x},\mathbf{z})$ in Eq.\eqref{eq:s49}, we get 
 \begin{align} K^{16}_{i}=&\sum_{x_1,z_1=0}^{7}
(-1)^{y_i^{x_1} +y_i^{z_1}}\sum_{x_2,z_2}(-1)^{x_2+z_2}\text{Tr}(\rho^a_{\mathbf{x}} \otimes \rho^c_{\mathbf{z}} \cdot B_i) \cdot \label{eq:4q3}
\end{align}
where $B_i$ is defined as 
\begin{align}
B_i=\sum_{b_1b_2b_3b_4}(-1)^{b_i} \pi_{b_1b_2b_3b_4}\label{eq:s54}
\end{align}
We define operators $A_i$ and $C_i$ analogous to previous sections as 
\begin{align}
A_i=\frac{1}{\omega^a_i}\sum_{x_1=0}^{7}(-1)^{y_i^{x_1}} \sum_{x_2=0}^{1}(-1)^{x_2} \rho^a_{x_1x_2}\notag\\ 
C_i=\frac{1}{\omega^c_i}\sum_{z_1=0}^{7}(-1)^{y_i^{z_1}} \sum_{z_2=0}^{1}(-1)^{z_2} \rho^c_{z_1z_2} \label{eq:s55}
\end{align}
Using the decomposition in Eq.\eqref{eq:s55}, we can write the quantum bound as
\begin{align}
\Delta^{16}_Q = \sum_{i=1}^{4} \sqrt{\big|\omega^a_{i} \omega_{i}^{c}\text{Tr}[(A_i\otimes C_i)  B_i]\big|}
\end{align}
Following the similar steps as Sec.\ref{sec:3}(See Eqs.\eqref{eq:s37}) and \eqref{eq:s38}), we have
\begin{align}
\Delta^Q \leq  
8\left(\sum_{i=1}^{4}(\omega^a_i)^2\right)^{\tfrac{1}{4}} 
\left(\sum_{i=1}^{4}(\omega^a_i)^2\right)^{\tfrac{1}{4}} 
\label{eq:B33}
\end{align}
We drive that $\forall i\,\, \max(\omega_i^a)=\max(\omega_i^c)=2 $(See Appendix~\ref{app:b4} for detailed derivation). Consequently,  
$
(\Delta^{16}_{Q})^{opt}=32 
.$
The optimal quantum bound is attained when 
\begin{align}B_i= A_i\otimes C_i  \,\,,\{A_i,A_j\}=0,\,\,,\{C_i,C_j\}=0 \,\,\forall{i\neq j}\label{eq:s59}
\end{align}
A detailed derivation of $(\Delta^Q_{16})^{opt}$, along with the corresponding maximization condition, is provided in Appendix~\ref{app:b4}.

Next, we demonstrate that, for the optimization, each measurement projector is maximally entangled across every \(1|3\) bipartition. Using the definition of $B_i$ in Eq.\eqref{eq:s54} we can write the measurement projectors as 
\begin{equation}
\pi_{b_1b_2b_3b_4}
=
\frac{1}{16}
\prod_{i=1}^{4}
\left(
\mathbb{I}_{16}
+
(-1)^{b_i} B_i
\right),
\qquad
b_i\in\{0,1\}.\label{eq:s60}
\end{equation}
It follows from Eq.~\eqref{eq:s55} that $\operatorname{Tr}(A_i)=\operatorname{Tr}(C_i)=0$. Moreover, the operators within each of the two sets $\{A_i\}$ and $\{C_i\}$ satisfy the anticommutation relations given in Eq.~\eqref{eq:s59}, namely, $\{A_i,A_j\}=0$ and $\{C_i,C_j\}=0$ for $i\neq j$. Using these anticommutation relations, we obtain, for pairwise distinct indices $i,j,k,l$,
$\operatorname{Tr}(A_iA_j)=\operatorname{Tr}(A_iA_jA_k)=\operatorname{Tr}(A_iA_jA_kA_l)=0$.
Analogous trace relations hold for products of the $C_i$ operators; see the detailed derivation in Appendix~\ref{app:b4}. Substituting the expressions for $B_i$ from Eq.~\eqref{eq:s55} into Eq.~\eqref{eq:s60} and using the above trace relations for products of the $A_i$ and $C_i$ operators, we obtain the reduced state of an arbitrary single-qubit subsystem by tracing out its complementary three-qubit subsystem. This gives
\begin{align}
\operatorname{Tr}_{\bar{i}}\!\left(\pi_{b_1b_2b_3b_4}\right)
=
\frac{\mathbb{I}_2}{2},
\qquad \forall\, i\in\{1,2,3,4\}.
\end{align}
Here, $\operatorname{Tr}_{\bar{i}}$ denotes the partial trace over all qubits except the $i^{th}$ qubit. Therefore, $(\Delta_Q^{16})^{opt}=32$, self-tests the measurement projector $\{\pi_{b_1b_2b_3b_4}\}$ to be maximally entangled across every \(1|3\) bipartition.
\section{Self testing of $2^m$ outcome joint measurement  for $2^m$ inputs per each sender \label{sec:7}}
In this section, we present a certification protocol for a joint measurement with $2^m$ outcomes, denoted by $\mathbf{b}=b_1b_2\cdots b_m \,\,\text{where}\,\, b_i \in \{0,1\} \,\text{for all}\, i=\{1,2,.,m\} $, performed by Bob. Alice and Charlie each randomly receive inputs, denoted by $\mathbf{x}=x_1x_2$ and $\mathbf{z}=z_1,z_2$, respectively, where $x_1,z_1\in\{0,1,2,\ldots,2^{m-1}-1\}$ and $x_2,z_2\in\{0,1\}$. We employ the following witness to certify the joint measurement
\begin{align}
\Delta^{2^m}=\sum_{i=1}^{m}\sqrt{|K_{i}^{2^{m}}|} 
\end{align}
where 
\begin{align}
&K_{i}^{2^{m}} =
\sum_{\substack{x_1,z_1=0}}^{2^{m-1}} 
(-1)^{y_i^{x_1} + y_i^{z_1}}\sum_{\mathbf{b},x_2,z_2}(-1)^{x_2+z_2+b_i}P(b|x_1x_2,z_1z_2) \label{eq:s64}
\end{align}
Here, each $y_i^{x_1}$ can take value either 0 or 1. To specify the value of $y_i^{x_1}$, we use the following encoding scheme.
Let $ D_m=\{0,1\}^m$ denote the set of all binary strings of length (m). We construct an ordered list $D_m=(y^0,y^1,\ldots,y^{2^m-1})$by first arranging the strings according to the number of 1s they contain, starting with strings containing no 1s and then increasing the number of 1s successively. If two strings contain the same number of 1s, they are arranged by comparing their bits from left to right, placing the string with 0 at the first position where they differ before the string with 1 at that position. For every string $y^x$, its bitwise complement is denoted by $\bar y^{x_1}=y^{x_1}\oplus_2(11\cdots1)$, where the complement is obtained by replacing every 0 with 1 and every 1 with 0. Under the ordering described above, the complementary string occurs at index $x_1'=2^{m-1}-x_1$. Consequently, $y^{x_1}\oplus_2 y^{x'_1}=11\cdots1$ and $x_1+x_1'=2^m-1$.
We select the first $2^{m-1}$ strings as the encoding set $ S_m=\{y^0,y^1,\ldots,y^{2^{m-1}-1}\}$. This construction gives $\mathcal S_3=\{000,001,010,100\}$ and $\mathcal S_4=\{0000,0001,0010,0100,1000,0011,0101,0110\}$, encoding schame used in Sec.\ref{sec:2},\ref{sec:3} and \ref{sec:4}. 
Finally, $y_i^x$ denotes the $i^{th}$ bit of the selected string $y^x$, where $i\in\{1,\ldots,m\}$ and $x_1\in\{0,1,\ldots,2^{m-1}-1\}.$ 

The classical causal model follows the decomposition in Eq.~\eqref{eq:s2}. Substituting this decomposition into Eq.~\eqref{eq:s64}, we have
\begin{align}
\Delta^{2^m}_{C} 
=
\sum_{i=1}^{m} \sqrt{\int d\lambda_1 
\left|
\sum_{x_1 =0}^{2^{m-1}-1}
(-1)^{y_i^{x_1}} \sum_{x_2=0}^{1} (-1)^{x_2}
P(\lambda_1|x_1x_2)
\right|}\notag\\ 
 \sqrt{\int d\lambda_2 \left|
\sum_{z_1=0}^{2^{m-1}-1}
(-1)^{y_i^{z_1}}\sum_{z_2=0}^{1} (-1)^{z_2}
P(\lambda_2|z_1z_2)
\right|}\label{eq:s65.a}  
\end{align}
Now using  $\sqrt{r_1r_2}+\sqrt{s_1s_2}+\sqrt{t_1t_2} \leq \sqrt{r_1+s_1+t_1}\cdot \sqrt{r_2+s_2+t_2}$ for all $r_1,r_2,s_1,s_2,t_1,t_2 \geq 0$
, we can write the classical bound as `
\begin{align}
\Delta^{2^m}_{C}  \leq 
\sqrt{\int d\lambda_1  F^a_m(\lambda_1)
}
\sqrt{\int d\lambda_2  F^c_m(\lambda_2)
}\label{eq:s66.a}
\end{align}
where
$F^a_m(\lambda_1)=\sum_{i=1}^{m}\bigg|\sum_{x_1=0}^{2^{m-1}-1}(-1)^{y^{x_1}_i}P(\lambda_1|x_1x_2)\bigg|$
and
$F^c_m(\lambda_2)=\sum_{i=1}^{m}\bigg|\sum_{x_1=0}^{2^{m-1}-1}(-1)^{y^{z_1}_i}P(\lambda_2|z_1z_2)\bigg|$.
We obtain
$(F^a_m)^{\max}=m\binom{m-1}{\left\lfloor\frac{m-1}{2}\right\rfloor}\big|\sum_{x_2}P(\lambda_1|x_1x_2)\big|$
and
$(F^c_m)^{\max}=m\binom{m-1}{\left\lfloor\frac{m-1}{2}\right\rfloor}\big|\sum_{x_2}P(\lambda_1|z_1z_2)\big|$(see derivation in Appendix~\ref{app:A2}).
This yields the classical causal upper bound
$\Delta^{2^m}_{C}\leq 2m\binom{m-1}{\left\lfloor\frac{m-1}{2}\right\rfloor}$.

In quantum case, Alice and Charlie's prepared states are denoted by $\rho^a_\mathbf{x} \equiv \rho^a_{x_1x_2}$ and $\rho^c_\mathbf{z}=\rho^c_{z_1z_2}$ respectively, where $x_1,z_1 \in \{0,1,.,.,2^{m-1}-1\}$ and $x_2,z_2 \in \{0,1\}.$ We assume that the dimension of both prepared systems is $2^{m/2}$. Bob then performs a $2^m$ outcome projective measurement denoted by $\{\pi_{b_1b_2...b_{m}}\}$ where $\sum_{b_1b_2...b_{m}=0}^{1}\pi_{b_1b_2...b_{m}}=\mathbb{I}_{2^m}$,\,$\pi_{b_1b_2b_3b_4} \geq 0$. We can write the joint probability distribution as  
$P(b_1b_2...b_{m}|\mathbf{x},\mathbf{z})=\text{Tr}(\rho^a_{
\mathbf{x}} \otimes \rho^c_{\mathbf{z}} \pi_{b_1b_2...b_{m}})$. By substituting the value of $P(b_1b_2...b_{m}|\mathbf{x},\mathbf{z})$ in Eq.\eqref{eq:s64}, we obtain 
 \begin{align} K^{2^m}_{i}=&\sum_{x_1,z_1=0}^{2^{m-1}-1}
(-1)^{y_i^{x_1} +y_i^{z_1}}\sum_{x_2,z_2=0}^{1}(-1)^{x_2+z_2}\text{Tr}(\rho^a_{\mathbf{x}} \otimes \rho^c_{\mathbf{z}} \cdot B_i) \cdot \label{eq:s65}
\end{align}
where $B_i$ is defined as 
\begin{align}
B_i=\sum_{b_1b_2..b_{m}}(-1)^{b_i} \pi_{b_1b_2..b_m}\label{eq:s66}
\end{align} 
By generalizing the derivation of the quantum bounds presented in Secs.~\ref{sec:3} and \ref{sec:4}, we obtain the optimal quantum bound $(\Delta^{2^m}_Q)^{\mathrm{opt}}=2^{m}\sqrt{m}$. The maximum is attained when $(\omega_i)^{\mathrm{opt}}B_i=A_i\otimes C_i$ along with $\{A_i,Aj\}=\{C_i.C_j\}=0 \,\forall \ i\neq j$, where the optimal values $\omega_i^{\mathrm{opt}}=\frac{2^{m}}{m}$ and the operators $A_i$ and $C_i$ are given below.
\begin{align}
A_i=\frac{1}{\omega^a_i}\sum_{x_1=0}^{2^{m-1}-1}(-1)^{y_i^{x_1}} \sum_{x_2=0}^{1}(-1)^{x_2} \rho^a_{x_1x_2}\notag\\ 
C_i=\frac{1}{\omega^c_i}\sum_{z_1=0}^{2^{m-1}-1}(-1)^{y_i^{z_1}} \sum_{z_2=0}^{1}(-1)^{z_2} \rho^c_{z_1z_2} \label{eq:s67}
\end{align}
Using the maximization condition, we show that, for the set
$\{A_1, A_2, \ldots, A_m\}$, every nonempty product of distinct
operators from this set is traceless i.e.,
$
\operatorname{Tr}\left(\prod_{i \in S} A_i\right) = 0,
\,
\forall\, S \subseteq \{1,2,\ldots,m\}.
$
The same property holds for the set
$\{C_1, C_2, \ldots, C_m\}$. Using these properties, we show that,
for even $m$, the measurement projectors
are maximally entangled across every $1|(m-1)$ bipartition. We provide the detailed derivation of quantum bound and maximization conditions in Appendix~\ref{app:b5}.

\section{Summary and Discussion}
In summary, we presented a SDI  protocol for certifying entangled basis measurement in a PM scenario featuring  two senders and one receiver. Each sender receives one of $2^m$ possible inputs and prepares a state accordingly, while the receiver performs a fixed $2^m$ outcome joint measurement on the composite system received from the senders. We construct suitable witnesses based on the number of inputs and derive the corresponding classical causal bounds. We demonstrate that the optimal quantum value of each witness exceeds the corresponding classical causal bound. Furthermore, the optimal quantum violation self-tests a class of $m$-qubit entangled-basis measurements whose eigenprojectors are maximally entangled across every $1|(m-1)$ bipartition for even $m$. In addition, we also consider the odd $m=3$ case, for which we obtain a weaker self-testing result that guarantees maximal entanglement across at least one bi-partition. A SDI protocol has been demonstrated to certify the $n$-qubit Greenberger-Horne-Zeilinger (GHZ) basis measurement in a multipartite PM scenario \cite{Debashis_2026_GHZ_mvmn-fzq7} which involves arbitrary parties and four inputs per sender. In contrast, we considered tri-partite scenario where the sender is restricted to two but allowing arbitrary number of inputs per sender.

An important direction for future work is to develop a self-testing protocol that certifies maximal entanglement across every $1|(m-1$ bipartition for both even and odd values of $m$, and  extend this framework to more general multipartite PM scenarios involving arbitrary numbers of parties and inputs. In recent times, various types of joint measurements have emerged as important resources for demonstrating different forms of nonlocality in quantum networks\cite{bilocal_vs_nonbilocal_PhysRevA.85.032119,entropy_21030325,Bilocal_EJM_PhysRevLett.126.220401}. Therefore, it would be interesting to explore their advantages in various quantum information processing tasks and to develop self-testing protocols for such joint quantum measurements within multipartite PM scenarios. The non-projective measurements~\cite{armin_2024_non_projective_BSM_PhysRevA.110.042206} have also been studied in both qubit systems and higher-dimensional settings. It is then worth exploring the extension of SDI self-testing protocols to non-projective measurements~\cite{Armin_Tavakoli_et_al._Self-testing_nonprojective_quantum_measurements_doi:10.1126/sciadv.aaw6664} within the multipartite PM scenario. 

\begin{widetext}

\appendix
\section{ Calculation of Classical Causal model bound }
\subsection{ Classical causal bound in case of two sender parties with 4 inputs per sender \label{app:appA1}}
Using Eq.~\eqref{eq:s6} from the main text, we can write
\begin{align}
\Delta^4_{C}=\sum_{i=1}^{2} \sqrt{\left|K_{i}\right|}
\le
\sum_{i=1}^{2} \sqrt{\int d\lambda_1 
\left|
\sum_{x_1x_2}
(-1)^{y^{x_{1}}_i+x_2}
P(\lambda_1|x_1x_2)
\right|} 
 \sqrt{\int d\lambda_2 \left|
\sum_{z_1z_2}
(-1)^{y^{z_1}_i+z_2}
P(\lambda_2|z_1z_2)
\right|}
\end{align}
To avoid cumbersome expression, we introduce notation 
$d_{x_1}^a
=
\sum_{x_2} (-1)^{x_2} P(\lambda_1|x_1x_2),
\quad
d^c_{z_1}
=
\sum_{z_2} (-1)^{z_2} P(\lambda_2|z_1z_2)$. Now using $\sqrt{r_1r_2}+\sqrt{s_1s_2} \leq \sqrt{r_1+s_1}\cdot \sqrt{r_2+s_2}$ for all $r_1,r_2,s_1,s_2 \geq 0$
we obtain
\begin{align}
\Delta^4_{C}
&\le
\sqrt{
\int d\lambda_1 \big(|d^a_0 + d^a_1| + |d^a_0 - d^a_1|\big)
}
\sqrt{
\int d\lambda_2 \big(|d^c_0 + d^c_1| + |d^c_0 - d^c_1|\big)
}
\end{align}
 
Using the bounds $|d^a_{x_1}|\leq \sum_{x_2} P(\lambda_1|x_1x_2)$ and $|d^c_{z_1}|\leq \sum_{z_2} P(\lambda_2|z_1z_2)$, together with the normalization conditions $\int d\lambda_1,P(\lambda_1|x_1x_2)=\int d\lambda_2,P(\lambda_2|c_1c_2)=1$, we obtain the classical causal upper bound $\Delta_C^4\leq 4$.

\subsection{Derivation of classical causal bound in case of $2^m$ outcome for receiver and $2^m$ inputs per sender  \label{app:A2}}
To derive the classical causal model bound $\Delta_C^{2^m}$ in Eq.~\eqref{eq:s65.a}, we first determine the optimal values of $F_m^a$ and $F_m^c$ (see Eq.~\eqref{eq:s66.a}). To this end, we introduce the following shorthand notation:
$d^a_{x_1}=\sum_{x_2}(-1)^{x_2}P(\lambda_1|x_1x_2)$ and
$d^c_{z_1}=\sum_{z_3}(-1)^{z_2}P(\lambda_2|z_1z_2)$.
We also define $M_{ix_1}=(-1)^{y^{x_1}_i}$ and $p=2^{m-1}$. Using this notation, we can write
\begin{align}
F^a_{m}=\sum_{i=1}^{m}\bigg|\sum_{x_1=0}^{p-1}M_{ix_{1}}d^a_{x_1}
\bigg|\end{align}
Using the identity
$
|t|=\max_{\epsilon\in\{-1,+1\}}\epsilon t,
$
which holds for any real-valued $t$, we have 
\begin{align}
 F^a_m=
\max_{\epsilon_1,\ldots,\epsilon_m\in\{-1,+1\}}
\sum_{i=1}^{m}\epsilon_i
\sum_{x_1=0}^{p-1}M_{ix_{1}}d_{x_1}.   \end{align}
The function $F_m^a$ is a sum of absolute values of linear combinations of the variables $d_{x_1}^a$. Therefore, $F_m^a$ is a convex, piecewise-linear function of these variables.  Since for all $x_1$ ,$-|\sum_{x_2} P(\lambda_1|x_1x_2)|\leq d^a_{x_1}\leq|\sum_{x_2}P(\lambda_1|x_1x_2)|$, we can write  
\begin{align}
    (F^a_m)^{\max}=
\max_{ d \leq G}
\max_{\boldsymbol\epsilon}
\sum_{i=1}^{m}\sum_{x_1=0}^{N-1}
\epsilon_i M_{ix_1}d_{x_1}.
\end{align}
where $d=(d^a_1,d^a_2,\ldots,d^a_{p-1})$, $\boldsymbol{\epsilon}=(\epsilon_1,\ldots,\epsilon_m)$, and $G=\left|\sum_{x_2}P(\lambda_1|x_1x_2)\right|$. For any real-valued $f$, we have $\max_{|d|\leq G}fG=G|f|$. Using this, we can write
 \begin{align}
    (F^a_m)^{\max}=
G
\max_{\boldsymbol\epsilon}
\sum_{x_1=0}^{N-1}\bigg|\sum_{i=1}^{m}
 M_{ix_1}\epsilon_i\bigg|
\end{align}
The expression $\sum_{i=1}^{m}M_{ix_1}\epsilon_i$ is the $x_1$-th component of the vector $M^{\mathsf T}\boldsymbol{\epsilon}$, where $M=[M_{ix_1}]$ is an $m\times 2^{m-1}$ matrix and $\boldsymbol{\epsilon}=(\epsilon_1,\ldots,\epsilon_m)^{\mathsf T}$. Therefore, taking the modulus of each component and summing over $x_1$ gives $\sum_{x_1=1}^{2^{m-1}}\left|\sum_{i=1}^{m}M_{ix_1}\epsilon_i\right|$. By construction, each coefficient $M_{ix_1}$ is either $+1$ or $-1$, as determined by the encoding scheme described in the main text. Hence, each modulus term is the absolute value of a linear combination of the $m$ variables $\epsilon_i$, with coefficients belonging to $\{-1,+1\}$.

The encoding scheme is constructed so that flipping the sign of any $\epsilon_i$ may change the individual modulus terms but only permutes their values. Consequently, the sum of all modulus terms remains invariant under any combination of sign flips. A more detailed illustration of this argument can be found in the Appendix of~\cite{Rahul_2021_PhysRevA.104.042217}, where a function employing a similar encoding scheme is optimized. Since there are $2^m$ possible sign assignments for $\boldsymbol{\epsilon}$, we may, without loss of generality, set $\epsilon_i=+1$ for every $i$.

We now count the contributions of the individual modulus terms. For a given $x_1$, let $j$ denote the number of terms $\epsilon_i$ that have coefficient $-1$ inside the modulus $\left|\sum_{i=1}^{m}M_{ix_1}\epsilon_i\right|$. Since the remaining $m-j$ terms have coefficient $+1$, setting every $\epsilon_i=+1$ gives a sum of $(m-j)-j=m-2j$. By choosing one representative from each pair of complementary encoding strings, we may take $j\leq\lfloor m/2\rfloor$. Therefore, the contribution of each modulus term is $m-2j$. For odd $m$, there are $\binom{m}{j}$ modulus terms containing exactly $j$ coefficients equal to $-1$, for $j=0,1,\ldots,\lfloor m/2\rfloor$. For even $m$, there are $\binom{m}{j}$ such modulus terms when $j<m/2$. However, when $j=m/2$, each encoding string and its bitwise complement contain the same number of negative coefficients. Thus, only half of these strings are selected as representatives, giving $\frac{1}{2}\binom{m}{m/2}$ modulus terms. Each of these contributes zero because it contains equal numbers of positive and negative coefficients.

Therefore, for both odd and even $m$, the sum of all modulus terms can be written as $\max_{\boldsymbol\epsilon}
\sum_{x_1=0}^{N-1}|\sum_{i=1}^{m}
 M_{ix_1}\epsilon_i|=\sum_{j=0}^{\lfloor m/2\rfloor}\binom{m}{j}(m-2j)$. Here, $\binom{m}{j}$ represents the number of modulus terms containing exactly $j$ coefficients of $\epsilon_i$ equal to $-1$, while $m-2j$ represents the corresponding sum inside each modulus when all $\epsilon_i=+1$. By using binomial-coefficient identity we can further write 
 \begin{align}
     \sum_{j=0}^{\lfloor m/2\rfloor}\binom{m}{j}(m-2j)=m\binom{m-1}{\left\lfloor\frac{m-1}{2}\right\rfloor}
 \end{align}
 This yields $(F^a_m)^{\max}=m\binom{m-1}{\left\lfloor\frac{m-1}{2}\right\rfloor}|\sum_{x_2}P(\lambda_1|x_1x_2)|$
, Similar calculation can carried out for $(F^c_m)^{max}$ and  we get $(F^c_m)^{max}=m\binom{m-1}{\left\lfloor\frac{m-1}{2}\right\rfloor}|\sum_{z_2}P(\lambda_1|z_1z_2)|$.  Since both $\int d\lambda_1 \left|\sum_{x_2} P(\lambda_1|x_1x_2)\right|$ and $\int d\lambda_2 \left|\sum_{z_2} P(\lambda_2|z_1z_2)\right|$ are bounded above by $2$, Eq.~\eqref{eq:s66.a} yields the classical causal bound $
\Delta_C^{2^m}\leq 2m\binom{m-1}{\left\lfloor\frac{m-1}{2}\right\rfloor}.
$
In particular, for $m=2$, $m=3$, and $m=4$, we obtain $\Delta_C^4\leq 4$, $\Delta_C^8\leq 12$, and $\Delta_C^{16}\leq 24$, respectively. 
\section{Derivation of optimal quantum bound}
\subsection{Quantum bound for the 8 input per sender and 4-outcome measurement for receiver  \label{app:appendixB_1} }
In this scenario, Alice and Charlie each prepare a single-qubit state and send it to Bob. Bob then performs a four-outcome measurement on the two-qubit system. We denote the quantum bound by $\Delta^8_{Q}$. Using  Eq.\eqref{eq:s34}, $K_i^8$ can be written as 
\begin{align}
K^8_i=Tr\left[\sum_{x,z =0}^{3}(-1)^{y_i^{x_1}}(-1)^{y_i^{z_1}} \sum_{x_2,z_2=0}^{1}(-1)^{{x_2}+{z_2}} (\rho^a_{x_1x_2} \otimes \rho^c_{z_1z_2})(B_i)\right]
\end{align}
In following we expand $K_i$ is terms of $\rho^a_{x_1x_2}$ and $\rho^c_{z_1z_2}$ for i=1,2,3
\begin{align}
K^8_1 = \text{Tr}
[(\rho^a_{00} - \rho^a_{01} + \rho^a_{10} - \rho^a_{11}
+ \rho^a_{20} - \rho^a_{21} - \rho^a_{30} + \rho^a_{31}) 
\otimes (\rho^c_{00} - \rho^c_{01} + \rho^c_{10} - \rho^c_{11}
+ \rho^c_{20} - \rho^c_{21} - \rho^c_{30} + \rho^c_{31})
\cdot B_1], \\
K^8_2 = \text{Tr}
[(\rho^a_{00} - \rho^a_{01} + \rho^a_{10} - \rho^a_{11}
- \rho^a_{20} + \rho^a_{21} + \rho^a_{30} - \rho^a_{31})
\otimes (\rho^c_{00} - \rho^c_{01} + \rho^c_{10} - \rho^c_{11}
- \rho^c_{20} + \rho^c_{21} + \rho^c_{30} - \rho^c_{31})
\cdot B_2], \\
K^8_3 = \text{Tr}
\Big[
(\rho^a_{00} - \rho^a_{01} - \rho^a_{10} + \rho^a_{11}
+ \rho^a_{20} - \rho^a_{21} + \rho^a_{30} - \rho^a_{31}) 
\otimes
(\rho^c_{00} - \rho^c_{01} - \rho^c_{10} + \rho^c_{11}
+ \rho^c_{20} - \rho^c_{21} + \rho^c_{30} - \rho^c_{31})
\cdot B_3
\Big].
\end{align}
where $B_i=\sum_{b_1,b_2=0}^{1} (-1)^{b_i} E_{b_1b_2}$
with $b_{i=1}=b_1$,$b_{i=2}=b_2$ and $b_{i=3}=b_1 \oplus b_2$\\
As presented in the main text maximum quantum bound can be written as 
 \begin{align}
\Delta^{8}_Q \leq 2\cdot \sqrt[4]{3((\omega^a_1)^2+(\omega^a_2)^2+(\omega^a_3)^2)}\cdot \sqrt[4]{3((\omega^c_1)^2+(\omega^c_2)^2+(\omega^c_3)^2)}
\end{align}
Using the definition of $\omega$ provided in Eq.\eqref{eq:s10.5}, we expand $\sum_{i=1}^{3}(\omega^a_i)^2$ as 
\begin{eqnarray}
\sum_{i=1}^{3}(\omega^a_i)^2
=\frac{1}{2}Tr\bigg[
3\big((\rho^a_{00})^2+(\rho^a_{01})^2+(\rho^a_{10})^2+(\rho^a_{11})^2
+(\rho^a_{20})^2+(\rho^a_{21})^2+(\rho^a_{30})^2+(\rho^a_{31})^2\big)
-3\{\rho^a_{00},\rho^a_{01}\}-3\{\rho^a_{10},\rho^a_{11}\}-3\{\rho^a_{20},\rho^a_{21}\}\notag\\-3\{\rho^a_{30},\rho^a_{31}\}
+\{\rho^a_{00}-\rho^a_{01},
\rho^a_{10}-\rho^a_{11}
+\rho^a_{20}-\rho^a_{21}
+\rho^a_{30}-\rho^a_{31}\}
-\{\rho^a_{10}-\rho^a_{11},
\rho^a_{20}-\rho^a_{21}
+\rho^a_{30}-\rho^a_{31}\}
-\{\rho^a_{20}-\rho^a_{21},\rho^a_{30}-\rho^a_{31}\}\bigg].
\end{eqnarray}
The above expression can be rewritten as
\begin{align}
\sum_{i=1}^{3}(\omega^a_i)^2
=
\frac{1}{2}Tr\![
4((\rho^a_{00})^2+(\rho^a_{01})^2+(\rho^a_{10})^2+(\rho^a_{11})^2
+(\rho^a_{20})^2+(\rho^a_{21})^2+(\rho^a_{30})^2+(\rho^a_{31})^2)\\
-2\{\rho^a_{00},\rho^a_{01}\}-2\{\rho^a_{10},\rho^a_{11}\}
-2\{\rho^a_{20},\rho^a_{21}\}-2\{\rho^a_{30},\rho^a_{31}\}
-(\delta^{a})^2
].
\end{align}
where \,\,$
\delta^a=\rho^a_{00}-\rho^a_{01}-\rho^a_{10}+\rho^a_{11}
-\rho^a_{20}+\rho^a_{21}-\rho^a_{30}+\rho^a_{31}.
$
For the maximum value,
\begin{align}
Tr
(\{\rho^a_{00}\rho^a_{01})=
Tr(\rho^a_{10}\rho^a_{11})=
Tr(\rho^a_{20}\rho^a_{21})=
Tr(\rho^a_{30},\rho^a_{31})=Tr(\delta^2)
=0
\end{align}
since $\delta^a$ is hermition operator, $ 
Tr[\delta^a)^2]=0\Longrightarrow \delta^a=0\Longrightarrow
\rho^a_{00}-\rho^a_{01}
-\rho^a_{10}+\rho^a_{11}
-\rho^a_{20}+\rho^a_{21}
-\rho^a_{30}+\rho^a_{31}
=0$. By doing pre and post-multiply by $\rho^a_{x_1x_2}$ for every $x_1$ and $x_2$, one will have 8 equations in terms of $\{ \rho_{ij}, \rho_{lm}\}$ where $i,j,l,m \in (0,1)$ 
Solving those leads to 
\begin{align}
Tr[(\rho^a_{00}-\rho^a_{01})(\rho^a_{10}-\rho^a_{11})]=
Tr[(\rho^a_{00}-\rho^a_{01})(\rho^a_{20}-\rho^a_{21})]=
Tr[(\rho^a_{00}-\rho^a_{01})(\rho^a_{30}-\rho^a_{31})]=\frac{2}{3},\notag\\
Tr[(\rho^a_{10}-\rho^a_{11})(\rho^a_{20}-\rho^a_{21}]=
Tr[(\rho^a_{10}-\rho^a_{11})(\rho^a_{30}-\rho^a_{31})]=
Tr[(\rho^a_{20}-\rho^a_{21})(\rho^a_{30}-\rho^a_{31})]=\frac{-2}{3}
\end{align}
The same relations hold for Charlie's prepared states:
\begin{align}
Tr[(\rho^c_{00}-\rho^c_{01})(\rho^c_{10}-\rho^c_{11})]=
Tr[(\rho^c_{00}-\rho^c_{01})(\rho^c_{20}-\rho^c_{21})]=
Tr[(\rho^c_{00}-\rho^c_{01})(\rho^c_{30}-\rho^c_{31})]=\frac{2}{3},\notag\\
Tr[(\rho^c_{10}-\rho^c_{11})(\rho^c_{20}-\rho^c_{21}]=
Tr[(\rho^c_{10}-\rho^c_{11})(\rho^c_{30}-\rho^c_{31})]=
Tr[(\rho^c_{20}-\rho^c_{21})(\rho^c_{30}-\rho^c_{31})]=\frac{-2}{3}
\end{align}
Below we provide an example of a set of prepared states and an entangled measurement that attain the optimal quantum value 
\begin{align}
&\rho_{0x_2}=\frac{1}{2}(\mathbb{I}_2+(-1)^{x_2}\frac{\sigma_x+\sigma_y+\sigma_z}{\sqrt{3}}) \,\, \forall x_2 \in\{0,1\} ,\quad \rho_{0z_2}=\frac{1}{2}(\mathbb{I}_2+(-1)^{z_2}\frac{\sigma_x+\sigma_y+\sigma_z}{\sqrt{3}}) \,\, \forall z_2 \in\{0,1\} ,\notag\\ 
&\rho_{1x_2}=\frac{1}{2}(\mathbb{I}_2+(-1)^{x_2}\frac{\sigma_x+\sigma_y-\sigma_z}{\sqrt{3}}) \,\, \forall x_2 \in\{0,1\}  , \quad \rho_{1z_2}=\frac{1}{2}(\mathbb{I}_2+(-1)^{z_2}\frac{\sigma_x+\sigma_y-\sigma_z}{\sqrt{3}}) \,\, \forall z_2 \in\{0,1\}  \notag\\
&\rho_{2x_3}=\frac{1}{2}(\mathbb{I}_2+(-1)^{x_2}\frac{\sigma_x-\sigma_y+\sigma_z}{\sqrt{3}}) \,\, \forall x_2 \in\{0,1\}, \quad \rho_{2z_3}=\frac{1}{2}(\mathbb{I}_2+(-1)^{z_2}\frac{\sigma_x-\sigma_y+\sigma_z}{\sqrt{3}}) \,\, \forall z_2 \in\{0,1\}\notag\\
&\rho_{3x_4}=\frac{1}{2}(\mathbb{I}_2+(-1)^{x_2}\frac{-\sigma_x+\sigma_y+\sigma_z}{\sqrt{3}}) \,\, \forall x_2 \in\{0,1\}, \quad \rho_{3z_4}=\frac{1}{2}(\mathbb{I}_2+(-1)^{z_2}\frac{-\sigma_x+\sigma_y+\sigma_z}{\sqrt{3}}) \,\, \forall z_2 \in\{0,1\}\notag\\
&\pi_{b_1b_2}=\frac{1}{4}\big(\mathbb{I}_2 \otimes \mathbb{I}_2+(-1)^{b_1}\sigma_{x} \otimes \sigma_{x}+(-1)^{b_2}\sigma_{y} \otimes \sigma_{y}+(-1)^{b_1+b_2}\sigma_{z} \otimes \sigma_{z}\big) \qquad \qquad \quad   \forall b_1,b_2 \in \{0,1\} \label{eq:b12.5}
\end{align}
\subsection{Derivation of optimal Quantum bound in case 8 inputs per sender and  8 outcome joint measurement \label{App:b3}}
Alice and Charlie each receive 8 inputs, denoted by $\mathbf{x}\equiv x_1x_2$ and $\mathbf{z}\equiv z_1z_2$, where $x_1,z_1\in\{0,1,2,3\}$ and $x_2,z_2\in\{0,1\}$. 
Bob performs a joint measurement on the received systems and produces 8 outcomes, denoted by $\{b_1b_2b_3\}$, where $b_i\in\{0,1\}$ for all $i\in\{1,2,3\}$. As presented in Sec.~\ref{sec:4} of the main text, we employ the following witness:
\begin{align}
\ \Delta^8=\sum_{i=1}^{3} \sqrt{\left|K_{i}^8\right|} 
\,\,\,\text{where}\,\, K^8_{i}=\sum_{x_1,z_1=0}^{3}
(-1)^{y_i^{x_1} +y_i^{z_1}}\sum_{x_2,z_2,b_1b_2,b_3}(-1)^{x_2+z_2+b_i}P(b_1b_2b_3|x_1x_2,z_1z_2) \label{eq:B13}
\end{align}
In the quantum case, Alice's and Charlie's preparations are denoted by $\rho^a_{\mathbf{x}}=\rho^a_{x_1x_2} \in \mathscr{L}(\mathscr{H}_a)$, $\dim(\mathscr{H}_a) =4$ and $\rho^c_{\mathbf{z}}=\rho^c_{z_1z_2} \in \mathscr{L}(\mathscr{H}_c)$, $\dim(\mathscr{H}_c) =2$, respectively. Bob's eight-outcome projective measurement is denoted by $\{\pi_{b_1b_2b_3}\}$, where $b_i\in\{0,1\}$ for all $i\in\{1,2,3\}$. For notational convenience, we define
For notation convenience, we define operators $\mathscr{A}_i \in \mathscr{L}({H}_A)$ and  $\mathscr{C}_i \in \mathscr{L}({H}_C)$ below
\begin{eqnarray}
  \mathscr{A}_1=   
\rho^a_{00} - \rho^a_{01} + \rho^a_{10} - \rho^a_{11}
+ \rho^a_{20} - \rho^a_{21} - \rho^a_{30} + \rho^a_{31} \,\,,  \mathscr{C}_1=\rho^a_{00} - \rho^a_{01} + \rho^a_{10} - \rho^a_{11}
+ \rho^a_{20} - \rho^a_{21} - \rho^a_{30} + \rho^a_{31}\notag\\
  \mathscr{A}_2=\rho^a_{00} - \rho^a_{01} + \rho^a_{10} - \rho^a_{11}
- \rho^a_{20} + \rho^a_{21} + \rho^a_{30} - \rho^a_{31}\,\,,  \mathscr{C}_2=\rho^a_{00} - \rho^a_{01} + \rho^a_{10} - \rho^a_{11}
- \rho^a_{20} + \rho^a_{21} + \rho^a_{30} - \rho^a_{31}\notag\\
  \mathscr{A}_3=\rho^a_{00} - \rho^a_{01} - \rho^a_{10} + \rho^a_{11}
+ \rho^a_{20} - \rho^a_{21} + \rho^a_{30} - \rho^a_{31}\,\,, \mathscr{C}_3= \rho^a_{00} - \rho^a_{01} - \rho^a_{10} + \rho^a_{11}
+ \rho^a_{20} - \rho^a_{21} + \rho^a_{30} - \rho^a_{31}\label{eq:b12}
\end{eqnarray}
In the defined notation, we can write the quantum bound as:
\begin{align}
\Delta^8_Q =\sum_{i=1}^{3} \sqrt{ \left|  Tr[( \mathscr{A}_i \otimes  \mathscr{C}_i) B_i]\right| } \,\quad
\text{where} \,\,B_i=\sum_{b_1b_2b_3}(-1)^{b_i} \pi_{b_1b_2b_3} \label{eq:b15}
\end{align}
By the Cauchy-Schwarz inequality for two operators $X,Y$, we have $Tr(XY) \leq \sqrt{Tr(X^2)}\sqrt{Tr(Y^2)}$. Equality holds when $X$ and $Y$ are linearly dependent. Using this, we can write the quantum bound as
\begin{align}
\Delta^8_Q \leq \sum_{i=1}^{3}\sqrt{\sqrt{Tr(( \mathscr{A}_i\otimes  \mathscr{C}_i)^2)}\sqrt{Tr(B_i)^2}}\label{eq:b16}
\end{align}
Equality holds when $\omega_i B_i=\mathscr{A}_i \otimes \mathscr{C}_i$. Since $(B_i)^2=I_8$, this implies $\omega_i=\sqrt{\frac{Tr(\ \mathscr{A}_i)^2}{4}}\sqrt{\frac{Tr( \mathscr{C}_i)^2}{2}}$. Using this, we can further write
\begin{align}
    \Delta^8_Q\leq \sum_{i=1}^{3}\sqrt{8 \omega^a_i\omega^c_i} ,\,\,\,\quad \text{where}\,\,\, \omega^a_i= \sqrt{\frac{Tr( \mathscr{A}_i)^2}{4}} , \omega^c_i =\sqrt{\frac{Tr( \mathscr{C}_i)^2}{2}}\label{eq:b17}
\end{align}
Since $\omega^a_i,\omega^c_i\geq 0$ for all $i$, we can apply the inequality to obtain
$
\sum_{i=1}^{m} \sqrt{\omega^a_i\omega^c_i}
\leq
\sqrt{\sum_{i=1}^{m}\omega^a_i}
\sqrt{\sum_{i=1}^{m}\omega^c_i}.
$
Furthermore,
$
\sum_{i=1}^{m}\omega^a_i
=
\sqrt{m^2\sum_{i=1}^{m}(\omega^a_i)^2}.
$
Combining these inequalities, we obtain
\begin{align}
\Delta^8_Q \leq 2\sqrt2\cdot\sqrt[4]{9 \bigg((\omega^a_1)^2+(\omega^a_2)^2+(\omega^a_3)^2 \bigg)  \bigg((\omega^c_1)^2+(\omega^c_2)^2+(\omega^c_3)^2\bigg)} \label{eq:b18}
\end{align}
Now, substituting the values of $\omega^a_i$ and $\omega^c_i$ into $\sum_{i=1}^3(\omega^c_i)^2$ and $\sum_{i=1}^{3}(\omega^a_i)^2$, respectively, we obtain
\begin{align}
\sum_{i=1}^{3}(\omega^c_i)^2
=
\frac{1}{2}Tr\!\bigg[
&4\big(\sum_{z_1=0}^{3}\sum_{z_2=0}^1(\rho^c_{z_1z_2})^2\big)
-2\{\rho^c_{00},\rho^c_{01}\}-2\{\rho^c_{10},\rho^c_{11}\}
-2\{\rho^c_{20},\rho^c_{21}\}-2\{\rho^c_{30},\rho^c_{31}\}
-(\delta^{c})^2
\bigg].\label{eq:b19}
\end{align}
where \,\,$
\delta^c=\rho^c_{00}-\rho^c_{01}-\rho^c_{10}+\rho^c_{11}
-\rho^c_{20}+\rho^c_{21}-\rho^c_{30}+\rho^c_{31}.
$
The maximum value of $\sum_{i=1}^{3}(\omega^c_i)^2$ is attained when 
\begin{align}
Tr
(\rho^c_{00}\rho^c_{01})=
Tr(\rho^c_{10}\rho^c_{11})=
Tr(\rho^c_{20}\rho^c_{21})=
Tr(\rho^c_{30},\rho^c_{31})=Tr[(\delta^c)^2]
=0 \quad \text{Together with}\,\,\, Tr\big[(\rho^c_{z_1z_2})^2\big]=1 \label{eq:b20}
\end{align}
For $\rho^c_{z_1z_2} \in \mathcal{L}(\mathbb{C}^2)$, all the conditions given in Eq.\eqref{eq:b20} are compatible with each other. We provide an explicit realization at the end of this section. Similarly, we can write $\sum_{i=1}^{3}(\omega^a_i)^2$ as
\begin{align}
\sum_{i=1}^{3}(\omega^a_i)^2
=
\frac{1}{4}Tr\!\bigg[
&4\big(\sum_{z_1=0}^{3}\sum_{z_2=0}^1(\rho^a_{z_1z_2})^2\big)
-2\{\rho^a_{00},\rho^a_{01}\}-2\{\rho^a_{10},\rho^a_{11}\}
-2\{\rho^a_{20},\rho^a_{21}\}-2\{\rho^a_{30},\rho^a_{31}\}
-(\delta^{a})^2
\bigg].\label{eq:b21}
\end{align}
where \,\,$
\delta^a=\rho^a_{00}-\rho^a_{01}-\rho^a_{10}+\rho^a_{11}
-\rho^a_{20}+\rho^a_{21}-\rho^a_{30}+\rho^a_{31}.
$
For $\rho^c_{z_1z_2} \in \mathcal{L}(\mathbb{C}^4)$, the condition $Tr[(\delta^a)^2]=0$ forces $Tr[(\rho^a_{x_1x_2})^2]=\frac{1}{2}$. Therefore, to maximize $\sum_{i=1}^{3}(\omega^a_i)^2$, we cannot simply set $Tr[(\delta^a)^2]=0$, since $\sum_{z_1=0}^{3}\sum_{z_2=0}^1Tr[(\rho^a_{z_1z_2})^2]$ and $Tr[(\delta^a)^2]$ are not independent with each other. We therefore provide an alternative way to derive the maximum.

It follows from the definition of $B_i$ in Eq.~\eqref{eq:b15} that $B_1$, $B_2$, and $B_3$ mutually commute. Consequently, the operators in each of the sets $\{ \mathscr{A}_1, \mathscr{A}_2, \mathscr{A}_3\}$ and $\{ \mathscr{C}_1, \mathscr{C}_2, \mathscr{C}_3\}$ must either mutually commute or mutually anticommute. In $\mathcal{L}(\mathbb{C}^2)$, there are no non-degenerate mutually commuting hermitian traceless operators. Thus, both the sets $\{ \mathscr{C}_1, \mathscr{C}_2, \mathscr{C}_3\}$ and $\{ \mathscr{A}_1, \mathscr{A}_2, \mathscr{A}_3\}$ must be mutually anticommuting.i.e. 
\begin{align}
\{\mathscr{A}_i,\mathscr{A}_j\}=\{\mathscr{C}_i,\mathscr{C}_j\}=0 \quad \forall i \neq j \label{eq:b21.5}
\end{align}
Equality holds on the right-hand sides of Eqs.\ \eqref{eq:b17} and \eqref{eq:b18} when $\omega^a_{1}=\omega^a_2=\omega^a_3=\omega^a$ and $\omega^c_{1}=\omega^c_2=\omega^c_3=\omega^c$. Using these conditions together with Eq.\ \eqref{eq:b12}, we can write
\begin{align}
&4 (\rho^a_{00}-\rho^a_{11})=\omega^a(A_1+A_2+A_3)+\delta^a \,\, ,\,\,4 (\rho^a_{10}-\rho^a_{11})=\omega^a(A_1+A_2-A_3)-\delta^a\notag\\
&4 (\rho^a_{20}-\rho^a_{21})=\omega^a(A_1-A_2+A_3)-\delta^a\,\, ,\,\,4 (\rho^a_{30}-\rho^a_{31})=\omega^a(- A_1+A_2+A_3)-\delta^a\label{eq:b22}
\end{align}
where $A_i=\frac{A_i}{\omega} \implies (A_i)^2=\mathbb{I}_4$.
Since all $\rho^a_{ij}$ are density matrices, the trace norm of $\rho^a_{i0}-\rho^a_{i1}$ is bounded by $||\rho^a_{i0}-\rho^a_{i1}||_1 \leq 2$. The trace norm of an operator $O$ is defined as
$
\|O\|_1 := \operatorname{Tr}\left(\sqrt{O^\dagger O}\right).
$
Using this, we can write
\begin{align}
\|4(\rho^a_{00}-\rho^a_{01})\|_1 =\|\omega^a(A_1+A_2+A_3)+\delta^a\|_1 \leq 8
\end{align}
 Consider the four operators
$X_1=\frac{A_1+A_2+A_3}{\sqrt{3}}$,
$X_2=\frac{A_1+A_2-A_3}{\sqrt{3}}$,
$X_3=\frac{A_1-A_2+A_3}{\sqrt{3}}$,
and
$X_4=\frac{-A_1+A_2+A_3}{\sqrt{3}}$.
Using $(A_i)^2=\mathbb{I}_4$ together with $\{A_i,A_j\}=0$ for $i\neq j$, it follows that each $X_i$ is unitary. We now use the following special case of Hölder's inequality:
$Tr(XY) \leq \|Y\|_1$,
where $X$ is a unitary operator and $Y$ is Hermitian. Equality holds when $[X,Y]=0$. 
Using this, we can write
 \begin{align}
&Tr[(X_1(\omega^a(  \,A_1+A_2+A_3)+\delta^a))]\leq \|\omega^a(\,A_1+A_2+A_3)+\delta^a\|_1 \leq 8\\
&Tr[(X_2(\omega^a( \,A_1+A_2-A_3)-\delta^a))]\leq \|\omega^a(\,A_1+A_2-A_3)-\delta^a\|_1 \leq 8\\
&Tr[(X_3(\omega^a(\,A_1-A_2+A_3)-\delta^a))]\leq \|\omega^a(\,A_1-A_2+A_3)-\delta^a\|_1 \leq 8\\
&Tr[(X_4(\omega^a(-A_1+A_2+A_3)-\delta^a))]\leq \|\omega^a(-A_1+A_2+A_3)-\delta^a\|_1 \leq 8
\end{align}
Since $X_1-X_2-X_3-X_4=0$, summing the four inequalities gives
\begin{align}
16 \sqrt{3}\omega^a \leq 32 \implies \omega^a \leq \frac{2}{\sqrt{3}}.
\end{align}
The upper bound $\omega^a=\frac{2}{\sqrt{3}}$ is attained when $\|\rho^a_{i0}-\rho^a_{i1}\|_1=2 \implies \rho^a_{i0}\rho^a_{i1}=0$ and $[X_i,\delta^a]=0\implies [A_i,\delta^a]=0 \, \forall\,\,i=1,2,3$. Substituting the corresponding values of $\omega^a_i$ and $\omega^c_i$ into Eq.\eqref{eq:b18}, we obtain $(\Delta_Q)^{opt}=8\sqrt{3}$.

Next, we derive the conditions for the measurement projectors $\{\pi_{b_1b_2b_3}\}$ to attain the maximal quantum value. Using the definition of $B_i$ in Eq.~\eqref{eq:b15}, the measurement projectors can be written as
\begin{align}
\pi_{b_1b_2b_3}
=
\frac{1}{8}
\prod_{i=1}^{3}
\left(
\mathbb{I}_{8}
+
(-1)^{b_i} B_i
\right),
\qquad
b_i\in\{0,1\}.
\end{align}
We have already established that, at the optimum, $\omega B_i=\mathscr{A}_i\otimes\mathscr{C}_i$ (see the discussion below Eq.~\eqref{eq:b16}).  From Eq.~\eqref{eq:b12}, it follows that $\operatorname{Tr}(\mathscr{A}_i)=\operatorname{Tr}(\mathscr{C}_i)=0$. Moreover, using the anticommutation relations in Eq.~\eqref{eq:b21.5}, we obtain $\operatorname{Tr}(\mathscr{A}_i\mathscr{A}_j)=\operatorname{Tr}(\mathscr{C}_i\mathscr{C}_j)=0$ for $i\neq j$.
Furthermore, for three mutually anticommuting Hermitian traceless operators $\mathscr{C}_i,\mathscr{C}_j,\mathscr{C}_k\in\mathcal{L}(\mathcal{H}_C)$, their product is proportional to $\mathbb{I}_2$ for pairwise distinct $i,j,k$. On the other hand, from the definition in Eq.~\eqref{eq:b15}, we have $\operatorname{Tr}(B_iB_jB_k)=0$ for pairwise distinct $i,j,k$. Using $\omega B_i=\mathscr{A}_i\otimes\mathscr{C}_i$, it therefore follows that $\operatorname{Tr}(\mathscr{A}_i\mathscr{A}_j\mathscr{A}_k)=0$.
Substituting the relation $\omega B_i=\mathscr{A}_i\otimes\mathscr{C}_i$ into Eq.~\eqref{eq:s60} and using the above trace relations for products of the $\mathscr{A}_i$'s and $\mathscr{C}_i$'s, we obtain, upon taking the partial trace over the complementary four-dimensional subsystem $\mathcal{H}_A$, on which the $\mathscr{A}_i$ operators act,
\begin{align}
\operatorname{Tr}_{\mathcal{H}_A}\!\left(\pi_{b_1b_2b_3}\right)
=
\frac{\mathbb{I}_2}{2}.
\end{align}
Thus, for every measurement outcome, the corresponding state has a maximally mixed single-qubit marginal. Consequently, each measurement state is maximally entangled across the $\mathcal{H}_A|\mathcal{H}_C$ bipartition, i.e., across the $12|3$ bipartition. Next we provide a explicit example of prepared state and measurement attianing maximum quantum value 

\begin{align}
\rho^a_{0x_2}
&=\frac14
\left[
I_4+\frac{(-1)^{x_2}}{\sqrt{3}}
\left(
\sigma_X\otimes\sigma_Y
+I_2\otimes\sigma_X
+I_2\otimes\sigma_Z
\right)
\right]\bigg[
I_4-(-1)^{x_2}
\left(\sigma_X\otimes I_2\right)\bigg], \quad \forall x_2 \in\{0,1\}\notag
\end{align}
\begin{align}
&\rho^a_{1x_2}
=\frac14
\left[
I_4+\frac{(-1)^{x_2}}{\sqrt{3}}
\left(
\sigma_X\otimes\sigma_Y
+I_2\otimes\sigma_X
-I_2\otimes\sigma_Z
\right)
\right]\bigg[
I_4-(-1)^{x_2}
\left(\sigma_X\otimes I_2\right)
\bigg], \quad \forall x_2 \in\{0,1\}\notag
\end{align}
\begin{align}
&\rho^a_{2x_2}
=\frac14
\left[
I_4+\frac{(-1)^{x_2}}{\sqrt{3}}
\left(
\sigma_X\otimes\sigma_Y
-I_2\otimes\sigma_X
+I_2\otimes\sigma_Z
\right)
\right]\bigg[
I_4-(-1)^{x_2}
\left(\sigma_X\otimes I_2\right)
\bigg], \quad \forall x_2 \in\{0,1\} \notag
\end{align}
\begin{align}
&\rho^a_{3x_2}
=\frac14
\left[
I_4+\frac{(-1)^{x_2}}{\sqrt{3}}
\left(
-\sigma_X\otimes\sigma_Y
+I_2\otimes\sigma_X
+I_2\otimes\sigma_Z
\right)
\right]\bigg[
I_4-(-1)^{x_2}
\left(\sigma_X\otimes I_2\right)
\bigg], \quad \forall x_2 \in\{0,1\}\notag \\
&\pi_{b_1b_2b_3}
=\frac{1}{8}
\left(
\mathbb{I}_{8}
+
(-1)^{b_1} \sigma_{x}\otimes \sigma_{y} \otimes \sigma_{x}+(-1)^{b_2}\mathbb{I}_2\otimes \sigma_{x} \otimes \sigma_{y}+ (-1)^{b_3}\mathbb{I}_2\otimes \sigma_{z} \otimes \sigma_{z}
\right)
\forall b_i\in\{0,1\}.
\end{align}
Charlie's optimal preparation $\rho^c_{z_1z_2}$ in this case can taken same as Appendix~\ref{app:appendixB_1}(see Eq.\eqref{eq:b12.5}). 
\subsection{Derivation of optimal Quantum bound and self testing conditions in case 16 inputs per sender and  16 outcome joint measurement \label{app:b4}}
Alice and Charlie each receive one of 16 possible input choices, denoted by $\mathbf{x}={x_1x_2}$ and $\mathbf{z}={z_1z_2}$, where $x_1,z_1\in\{0,1,2,3,4,5,6,7\}$ and $x_2,z_2\in\{0,1\}$. Based on their inputs, Alice and Charlie prepare two-qubit states, denoted by $\rho^a_{\mathbf{x}}=\rho^a_{x_1x_2}$ and $\rho^c_{\mathbf{c}}=\rho^c_{z_1z_2}$, respectively, and send them to Bob. Bob then performs a joint projective measurement with 16 possible outcomes, denoted by ${\pi_{b_1b_2b_3b_4}}$, where $b_i\in\{0,1\}$ for all $i\in\{1,2,3,4\}$.
We optimize the expression $\sum_{i=1}^{4} \sqrt{\left|K_{i}^{16}\right|}$ and denote its quantum value by $\Delta^{16}_{Q}$. Therefore, from Eq.~\eqref{eq:s47}, we can write
\begin{align}\Delta^{16}_Q=\sum_{i=1}^{4} \sqrt{\left|K_{i}^{16}\right| } \label{eq:B12}
\end{align}
Using born rule, we can write  $K_i^{16}$ as  
 \begin{align}
 K^{16}_{i}=\sum_{x_1,z_1=0}^{7} 
(-1)^{y_i^{x_1} +y_i^{z_1}}\sum_{x_2,z_2}(-1)^{x_2+z_2}Tr(\rho^a_{x_1x_2}\otimes \rho^c_{z_1z_2}\cdot B_i)\,\, \text{where}\, B_i=\sum_{b_1b_2b_3b_4}(-1)^{b_i}\pi_{b_1b_2b_3b_4}\label{eq:B14}
\end{align}
For notational convenience, we define operators $ \mathscr{A}_i$ and $ \mathscr{C}_i$ as follows 
\begin{align}
\mathscr{A}_i=\sum_{x_1=0}^{7}(-1)^{y_i^{x_1}} \sum_{x_2=0}^{1}(-1)^{x_2} \rho_{x_1x_2}\,\quad , \,\,
\mathscr{C}_i=\sum_{z_1=0}^{7}(-1)^{y_i^{z_1}} \sum_{z_2=0}^{1}(-1)^{z_2} \rho_{z_1z_2}
\label{eq:B_4}
\end{align}
 By the Cauchy-Schwarz inequality for two operators $X,Y$, we have $Tr(XY) \leq \sqrt{Tr(X^2)}\sqrt{Tr(Y^2)}$. Equality holds when $X$ and $Y$ are linearly dependent. Using this, we can write the quantum bound as
\begin{align}
\Delta^{16}_Q \leq \sum_{i=1}^{4}\sqrt{\bigg|\sqrt{Tr(( \mathscr{A}_i\otimes  \mathscr{C}_i)^2)}\sqrt{Tr[(B_i)^2]}\bigg|}\label{eq:b35}
\end{align}
Equality holds when $\omega_i B_i=A_i \otimes C_i$. Since $(B_i)^2=I_8$, this implies $\omega_i=\sqrt{\frac{Tr( \mathscr{A}_i)^2}{4}}\sqrt{\frac{Tr( \mathscr{C}_i)^2}{4}}$. Using this, we can further write
\begin{align}
\sum_{i=1}^{4} \sqrt{\left|K_{i}^{16}\right|} \leq 4\sum_{i=1}^{4}\sqrt{|\omega^a_i\omega^c_i|}\quad\text{where}\,\,\, \omega^a_i= \sqrt{\frac{Tr( \mathscr{A}_i)^2}{4}} , \omega^c_i =\sqrt{\frac{Tr( \mathscr{C}_i)^2}{4}}\label{eq:B32}
\end{align}
Using Inequality 1: $\sum_{i=1}^{4}\sqrt{r_i s_i}\le \sqrt{\sum_{i=1}^{4} r_i}\,\sqrt{\sum_{i=1}^{4} s_i}$ for $r_i,s_i\ge 0$, and Inequality 2: $r_1+\cdots+r_4\le \sqrt{4(r_1^2+r_2^2+r_3^2+r_4^2)}$, the expression simplifies to
\begin{align}
\Delta^{16}_Q \leq  
8\left(\sum_{i=1}^{4}(\omega^a_i)^2\right)^{\tfrac{1}{4}} 
\left(\sum_{i=1}^{4}(\omega^a_i)^2\right)^{\tfrac{1}{4}} 
\label{eq:B33}
\end{align}
 Using the definition of $\omega^a_i$ and $\omega^c_i$(see Eq.\eqref{eq:B32}), we can write 
\begin{align}
4\sum_{i=1}^{4}(\omega^a_i)^2=\sum_{i=1}^{4} \bigg[ Tr\bigg (\sum_{x_1=0}^{7} \sum_{x_2=0}^{1}
(-1)^{y_i^{x_1}}(-1)^{x_2}\rho^a_{x
_1x_2}\bigg)^2\bigg] \quad ,\,\,
4\sum_{i=1}^{4}(\omega^c_i)^2=\sum_{i=1}^{4}  \bigg(\sum_{z=0}^{7}\sum_{z_4=0}^{1}
(-1)^{y_i^{z_1}} (-1)^{z_2}\rho^c_{z_1z_2}\bigg)^2\label{eq:B_5}
\end{align}
After expanding the summation over $x_1$, $x_2$, and $i$, and fixing the value of $y_i^{x_1}\in\{0,1\}$ according to the encoding scheme defined in the main text, we have
\begin{align}
4\sum_{i=1}^{4}(\omega^a_i)^2=\sum_{x_1=0}^{7} \sum_{x_2=0}^{1}  Tr\bigg(8\cdot (\rho^a_{x_1x_2})^2- \rho^a_{x_10}\rho^a_{x_11}-(\delta^a_1)^2-(\delta^a_2)^2-(\delta^a_3)^2-(\delta^a_4)^2\bigg) \label{eq:B_19}
\end{align}
where $\delta^a_1,\delta^a_2,\delta^a_3$ and $\delta^a_4$ are given as 
\begin{align}
\delta^a_1=\sum_{x_2=0}^{1}(-1)^{x_2} [\rho^a_{0x_2}-\rho^a_{1x_2}-\rho^a_{2x_2}-\rho^a_{3x_2}+\rho^a_{4x_2}+\rho^a_{5x_2}+\rho^a_{6x_2}+\rho^a_{7x_2}]\notag\\
\delta^a_2=\sum_{x_2=0}^{1}(-1)^{x_2} [\rho^a_{0x_2}-\rho^a_{1x_2}-\rho^a_{2x_2}+\rho^a_{3x_2}-\rho^a_{4x_2}+\rho^a_{5x_2}-\rho^a_{6x_2}-\rho^a_{7x_2}]\notag\\
\delta^a_3=\sum_{x_2=0}^{1}(-1)^{x_2} [\rho^a_{0x_2}-\rho^a_{1x_2}+\rho^a_{2x_2}-\rho^a_{3x_2}-\rho^a_{4x_2}-\rho^a_{5x_2}+\rho^a_{6x_2}-\rho^a_{7x_2}]\notag\\
 \delta^a_4=\sum_{x_2=0}^{1}(-1)^{x_2} [\rho^a_{0x_2}+\rho^a_{1x_2}-\rho^a_{2x_2}-\rho^a_{3x_2}-\rho^a_{4x_2}-\rho^a_{5x_2}-\rho^a_{6x_2}+\rho^a_{7x_2}]\label{eq:B_8}
 \end{align}
We define the scaled operators $A_i=\frac{ \mathscr{A}_i}{\omega^a_i}$. It follows directly from Eq.~\eqref{eq:B14} that $B_1,B_2,B_3,B_4$ form a set of mutually commuting operators. Therefore, each of the sets ${A_i}$ and ${C_i}$ must consist either of mutually commuting or mutually anticommuting operators. Given that the dimension of the prepared state $\rho^a_{x_1x_2}$ is four, and that in $d=4$ it is not possible to have four mutually commuting Hermitian, traceless operators, the set $(A_1,A_2,A_3,A_4)$ must therefore be mutually anticommuting. Similarly, the set ${C_i}$ also forms a mutually anticommuting set. Hence,
\begin{align}
\{A_i,A_j\}=\{C_i,C_j\}=0 \label{eq:B24}
\ \text{for all} \ i\neq j
\end{align}
Following the same approach as in Appendix~\ref{App:b3}, we define eight operators $X_1=\frac{A_1+A_2+A_3+A_4}{2}$, $X_2=\frac{A_1+A_2+A_3-A_4}{2}$,$X_3=\frac{A_1+A_2-A_3+A_4}{2}$,$X_4=\frac{A_1-A_2+A_3+A_4}{2}$,$X_5=\frac{-A_1+A_2+A_3+A_4}{2}$,$X_6=\frac{A_1+A_2-A_3-A_4}{2}$,$X_7=\frac{A_1-A_2+A_3-A_4}{2}$,$X_8=\frac{A_1-A_2-A_3+A_4}{2}$.The mutual anticommutation among the elements of the set ${\mathscr{A}i}$, together with $(\mathscr{A}i)^2=\mathbb{I}_4$, ensures that all $X_i$'s are unitary operators. Equality holds on the right-hand sides of Eqs.~\eqref{eq:B32} and \eqref{eq:B33} when $\omega^a{i}=\omega^a{j}, \forall , i,j$ and $\omega^c{i}=\omega^c_{j},\forall ,i,j$. By following the same steps as in Eqs.~\eqref{eq:b20}, \eqref{eq:b21}, and \eqref{eq:b22}, we can write each $8(\rho^a_{i0}-\rho^a_{i1})$ as a function of $A_i,\omega^a,\delta^a_i$, where $i=1,2,3,4$. Next, we use Hölder's inequality,
$Tr(XY) \leq |Y|_1$,
for a unitary operator $X$ and a Hermitian operator $Y$, and obtain the following inequalities
\begin{align}
&Tr[(X_1(\omega^a(  \,A_1+A_2+A_3+A_4)+\delta^a_{1}+\delta^a_2+\delta^a_3+\delta^a_4)] \leq 16\\
&Tr[(X_2(\omega^a(  \,A_1+A_2+A_3-A_4)-\delta^a_{1}-\delta^a_2-\delta^a_3+\delta^a_4)] \leq 16\\
&Tr[(X_3(\omega^a(  \,A_1+A_2-A_3+A_4)-\delta^a_{1}-\delta^a_2+\delta^a_3-\delta^a_4)] \leq 16\\
&Tr[(X_4(\omega^a(  \,A_1- A_2+A_3+A_4)-\delta^a_{1}+\delta^a_2-\delta^a_3-\delta^a_4)] \leq 16\\
&Tr[(X_5(\omega^a(-A_1+A_2+A_3+A_4)+\delta^a_{1}-\delta^a_2-\delta^a_3-\delta^a_4)] \leq 16\\
&Tr[(X_6(\omega^a(  \,A_1+A_2-A_3-A_4)+\delta^a_{1}+\delta^a_2-\delta^a_3-\delta^a_4)] \leq 16\\
&Tr[(X_7(\omega^a(  \,A_1-A_2+A_3-A_4)+\delta^a_{1}-\delta^a_2+\delta^a_3-\delta^a_4)] \leq 16\\
&Tr[(X_8(\omega^a(  \,A_1-A_2-A_3+A_4)+\delta^a_{1}-\delta^a_2-\delta^a_3+\delta^a_4)] \leq 16\\\notag
\end{align}
Upon summing the eight inequalities, all the terms involving $Tr(X_i\delta^a_j)$ cancel each other. Therefore, we obtain
\begin{align}
64 \omega^a \leq 128 \implies \omega^a \leq 2
\end{align}
Similar calculation can be carried out for $\omega^c\leq2$. This implies $\max \big(\sum_{i=1}^{4}(\omega^a_i)^2\big)=\max \big(\sum_{i=1}^{4}(\omega^c_i)^2\big) =16$. Substituting this into Eq.\eqref{eq:B33}, we obtain $(\Delta^{16}_{Q})^{opt}=32.$
Next, using the optimization condition derived above, we show that all the measurement projectors must be maximally entangled. 
To demonstrate this, let us consider a set \( G \) whose elements are all possible matrix products of \( B_1, B_2, B_3, \) and \( B_4 \). This set forms a group under matrix multiplication. Since \( B_1, B_2, B_3, \) and \( B_4 \) mutually commute, all elements of \( G \) also commute with each other. The set \( G \) contains 16 unique elements in total, including the identity, and its generating set is 
$
S = \{ B_1, B_2, B_3, B_4 \}.
$Each element of $ S $ can be expressed as a linear combination of projectors with coefficients $ +1 $ or $-1$. Consequently, every element of the group \( G \) can also be written as a linear combination of these projectors with weights \( \pm 1 \). This holds because all cross terms vanish due to the orthogonality of the projectors.

More formally, the 16 projectors of the joint measurement are linearly related to the 16 elements of \( G \) through a transformation matrix \( M \). Since the projectors are mutually orthogonal, each row of \( M \) contains an equal number of \( +1 \) and \( -1 \) entries, except for the row corresponding to the identity element, which consists entirely of \( +1 \)s. All rows of \( M \) are mutually orthogonal, implying that \( M \) is an orthogonal and hence invertible matrix. This means each projector can also be expressed as a linear combination of the elements of \( G \).

Therefore, taking the partial trace of any projector over three qubits is equivalent to taking the partial trace of each element of \( G \) over three qubits and then forming the same linear combination. Using this, we now show that the partial trace of every non-identity element of \( G \) over any three qubits is zero.

We present the calculation of above claim 
Let us  first consider the partial trace of the generating set S of group G, each  element of the set is 
\begin{align}
B_i=A_i\otimes C_i=\sum_{x_1,z_1=0}^{7}(-1)^{y_i^{x_1}} \cdot (-1)^{y_i^{z_1}} \sum_{x_2,z_2}(-1)^{x_2+z_2} \frac{\rho^a_{x_1x_2}}{\omega^a_i} \otimes \frac{\rho^c_{z_1z_2}}{\omega^c_i}
\end{align}
since $Tr\bigg(\sum_{x_1=0}^{7}(-1)^{y_i^{x_1}} \sum_{x_2}(-1)^{x_2} \frac{\rho^a_{x_1x_2}}{\omega^a_i}\bigg)=0$.
So, the partial trace of $B_i$ over any three qubits will be zero.
Next, we consider elements of 
G, which are products of two elements from the generating set
$B_iB_j=A_iA_j \otimes C_iC_j$ , since $\{A_i,A_j\}=0 \, \forall i\neq j$ we can write  $Tr(A_iA_j)=-Tr(A_jA_i)$ futhermore $Tr(A_iA_j)=0$, similar calculation can carried out for $C_iC_j$.
Next, we consider the elements of $G$, which are products of three elements from the generating set.
To show that the partial trace of \( B_1 B_2 B_3 \) over any three qubits is zero, it suffices to show that \( \mathrm{Tr}(A_1 A_2 A_3) = 0 \). We prove this by contradiction: assume that there exists a set \( A_1, A_2, A_3, A_4 \) of four mutually anti-commuting dichotomic traceless operators such that 
 \begin{align}
 A_1A_2A_3 = \beta  I +\gamma \cdot\Sigma\label{eq:B_14}
 \end{align}
 where $\Sigma$ is some traceless operator and $\beta$ and $\gamma$ are some coefficients, Since $A_1,A_2,A_3,A_4 $ mutually anti-commute with each other , we can write  
 \begin{align}
 \{A_4A_1A_2A_3,A_4\}=0 \label{eq:B_15}
 \end{align}
Pre-multiplying both side in  \eqref{eq:B_14} by $A_4$, we have
 \begin{align}
 A_4A_1A_2A_3=\beta  A_4+\gamma  A_4 \Sigma 
 \end{align}
 The above equation is compatible with Eq.~\eqref{eq:B_15} only when \( \beta = 0 \). This implies $\text{Tr}(A_1A_2
 A_3)=0$. In general for any set of four mutually anti-commuting dichotomic traceless Hermitian operators, \( \mathrm{Tr}(A_i A_j A_k) \) is always zero. This implies that the partial trace of \( B_i B_j B_k \) over any three-qubit subsystem is zero for all \( i, j, k \in \{1,2,3,4\} \). \\Next, we show that the partial trace of \( B_1 B_2 B_3 B_4 \) over any three qubits is also zero. To do this, we again use Eq.~\eqref{eq:B_15} and show that \( \mathrm{Tr}(A_4 A_1 A_2 A_3) = 0 \). From Eq.~\eqref{eq:B_15} we have \( A_4 \cdot T = - T \cdot A_4 \), where \( T = A_4 A_1 A_2 A_3 \). Then \( \text{Tr}(T) = -\text{Tr}(A_4 T A_4) = -\text{Tr}(T) \), which implies \( \mathrm{Tr}(T) = 0 \). Therefore, the partial trace of \( B_4 B_1 B_2 B_3 \) over any three qubits is also zero. Hence the only  term $\mathbb{I}_2 \otimes \mathbb{I}_2 \otimes \mathbb{I}_2 \otimes \mathbb{I}_2$ will contribute in the expression of reduced single qubit marginals i.e.,$\operatorname{Tr}_{\bar{i}}\!\left(\pi_{b_1b_2b_3b_4}\right)
=
\frac{\mathbb{I}_2}{2},
\,\, \forall\, i\in\{1,2,3,4\}$, Here, $\operatorname{Tr}_{\bar{i}}$ denotes the partial trace over all qubits except the $i$-th qubit, leaving an operator acting on subsystem $i$.
\subsection{Derivation of optimal Quantum bound and for $2^m$ outcome joint measurement and $2^m$ inputs per sender \label{app:b5}}
As described in the main text, we can write the quantum bound in this case as 
\begin{align}
\Delta^{2^m}=\sum_{i=1}^{m}\sqrt{|K_{i}^{2^{m}}|} \,\,\,\text{where}  \,\,K^{2^m}_{i}=&\sum_{x_1,z_1=0}^{2^{m-1}-1}
(-1)^{y_i^{x_1} +y_i^{z_1}}\sum_{x_2,z_2=0}^{1}(-1)^{x_2+z_2}\text{Tr}(\rho^a_{\mathbf{x}} \otimes \rho^c_{\mathbf{z}} \cdot B_i) \cdot \label{eq:s65}\,\,\,,B_i=\sum_{b_1b_2..b_{m}}(-1)^{b_i} \pi_{b_1b_2..b_m}
\end{align}
Analogous to  Appendix.~\ref{app:b4}, we define the operators $\mathscr{A}_i$ and $\mathscr{C}_i$ as
\begin{align}
\mathscr{A}_i=\sum_{x_1=0}^{7}(-1)^{y_i^{x_1}} \sum_{x_2=0}^{1}(-1)^{x_2} \rho_{x_1x_2}\,\quad , \,\,
\mathscr{C}_i=\sum_{z_1=0}^{7}(-1)^{y_i^{z_1}} \sum_{z_2=0}^{1}(-1)^{z_2} \rho_{z_1z_2}
\label{eq:B59}
\end{align}
Following the similar steps as in Eqs.~\eqref{eq:b35}, \eqref{eq:B32}, and
\eqref{eq:B33}, we use the following Identities $
\sum_{i=1}^{m} \sqrt{\omega^a_i\omega^c_i}
\leq
\sqrt{\sum_{i=1}^{m}\omega^a_i}
\sqrt{\sum_{i=1}^{m}\omega^c_i}.
$
and  
$
\sum_{i=1}^{m}\omega^a_i
=
\sqrt{m^2\sum_{i=1}^{m}(\omega^a_i)^2}.
$ Thus, we obtain the quantum bound as
\begin{align}
\Delta^{2^m}_Q \leq  
2^{m/2}\sqrt{m}\left(\sum_{i=1}^{m}(\omega^a_i)^2\right)^{\tfrac{1}{4}} 
\left(\sum_{i=1}^{m}(\omega^a_i)^2\right)^{\tfrac{1}{4}} 
\label{eq:B58}
\end{align}
Using the definition of $\omega^a_i$ given in Eq.~\eqref{eq:B32}, we can write
\begin{align}
2^{m/2}\sum_{i=1}^{m}(\omega^a_i)^2=\sum_{x_1=0}^{2^{m-1}-1} \sum_{x_2=0}^{1}  Tr\bigg(2^{m-1}\cdot (\rho^a_{x_1x_2})^2- \rho^a_{x_10}\rho^a_{x_11}-\sum_{l=1}^{2^{m-1}-m}(\delta^a_j)^2\bigg) \label{eq:B_19}
\end{align}
where $\delta^a_j=\sum_{x_1=0}^{2^{m-1}-1} (-1)^{l_{j}.y^{x_1}} (\rho^a_{x_{1}0}-\rho^a_{x_{1}1}).$ Here, $l_j$ denotes an $m$-bit binary string. To determine the possible values of $l_j$, consider the set $\mathcal{L}=\{0,1\}^m$, which contains all binary strings of length $m$. We define a subset $L\subseteq\mathcal{L}$ containing only those strings that satisfy the following two conditions. First, each string must contain at least two $1$s, i.e.,
$\sum_{r=1}^{m}(l_j)_r\geq 2$. Second, the number of $1$s in each selected string must be odd. Equivalently, $\sum_{r=1}^{m}(l_j)_r$ must not be an even integer. Thus, we define
$L_m=\{l_j\in\mathcal{L}:\sum_{r=1}^{m}l_r\geq2,\ \sum_{r=1}^{m}l_r\text{ is odd}\}$. The selected strings are denoted by $l_j$, where $j=1,2,\ldots,2^{m-1}-m.$ \\
following the similar approach as Appendix~\ref{app:b4}. we write 
\begin{align}
2^{m-1} (\rho^a_{x_10}-\rho^a_{x_21})=\sum_{i=1}^{m}(-1)^{y_i^{x_1}}\omega_i A_i+\sum_{j=1}^{2^{m-1}-m} (-1)^{l_j\cdot y^{x_1}} \delta^a_j
\end{align}
where $A_i$ is a scaled operator defined as $A_i=\frac{\mathscr{A}_i}{\omega^a_i}$. Next, we define the operators $X_{x_1}=\frac{1}{\sqrt{m}}\sum_{i=1}^{m}(-1)^{y_i^{x_1}}A_i$. Since $\|\rho^a_{i0}-\rho^a_{i1}\|_1\leq 2$, we apply Hölder's inequality, $|\operatorname{Tr}(XY)|\leq \|Y\|_1$, where $X$ is a unitary operator and $Y$ is a Hermitian operator, to obtain the following
\begin{align}
&Tr\big[(X_{x_1}\big(\sum_{i=1}^{m}(-1)^{y_i^{x_1}}\omega_i A_i+\sum_{j=1}^{2^{m-1}-m} (-1)^{l_j\cdot y^{x_1}} \delta^a_j\big)\big] \leq 2^{m-1}||\rho^a_{x_10}-\rho^a_{x_11}||_1
\end{align}
It follows from the construction of $X_{x_1}$ that $\sum_{x_1}(-1)^{l_j\cdot y^{x_1}}X_{x_1}=0$ for all $j$. Therefore,
\begin{align}
&\sum_{x_1=0}^{2^{m-1}-m}Tr\big[(X_{x_1}\big(\sum_{i=1}^{m}(-1)^{y_i^{x_1}}\omega_i A_i+\sum_{j=1}^{2^{m-1}-m} (-1)^{l_j\cdot y^{x_1}} \delta^a_j\big)\big]=2^{m-1}2^{m/2}\sqrt{m} \omega^a\leq 2^{m-1}2^{m-1}||\rho^a_{x_10}-\rho^a_{x_11}||_1
\end{align}
Therefore, we obtain $\omega^a_i \leq \frac{2^{m/2}}{\sqrt{m}}$. A similar calculation yields $\omega^c_i \leq  \frac{2^{m/2}}{\sqrt{m}} \forall i=1,2,.,m$. This implies that $\max\big(\sum_{i=1}^{m}(\omega^a_i)^2\big)=\max\big(\sum_{i=1}^{m}(\omega^c_i)^2\big)=2^m$. Substituting these results into Eq.~\eqref{eq:B33}, we obtain $(\Delta^{2^m}_{Q})^{\mathrm{opt}}=2^m\sqrt{m}$.
\end{widetext}
\bibliography{bib}

@article{J.S.BellPhysicsPhysiqueFizika.1.195,
  title = {On the Einstein Podolsky Rosen paradox},
  author = {Bell, J. S.},
  journal = {Physics Physique Fizika},
  volume = {1},
  issue = {3},
  pages = {195--200},
  numpages = {6},
  year = {1964},
  month = {Nov},
  publisher = {American Physical Society},
  doi = {10.1103/PhysicsPhysiqueFizika.1.195},
  url = {https://link.aps.org/doi/10.1103/PhysicsPhysiqueFizika.1.195}
}

@article{entanglement_swapping_1993PhysRevLett.71.4287,
  title = {``Event-ready-detectors'' Bell experiment via entanglement swapping},
  author = {\ifmmode \dot{Z}\else \.{Z}\fi{}ukowski, M. and Zeilinger, A. and Horne, M. A. and Ekert, A. K.},
  journal = {Phys. Rev. Lett.},
  volume = {71},
  issue = {26},
  pages = {4287--4290},
  numpages = {0},
  year = {1993},
  month = {Dec},
  publisher = {American Physical Society},
  doi = {10.1103/PhysRevLett.71.4287},
  url = {https://link.aps.org/doi/10.1103/PhysRevLett.71.4287}
}

@article{bilocal_vs_nonbilocal_PhysRevA.85.032119,
  title = {Bilocal versus nonbilocal correlations in entanglement-swapping experiments},
  author = {Branciard, Cyril and Rosset, Denis and Gisin, Nicolas and Pironio, Stefano},
  journal = {Phys. Rev. A},
  volume = {85},
  issue = {3},
  pages = {032119},
  numpages = {21},
  year = {2012},
  month = {Mar},
  publisher = {American Physical Society},
  doi = {10.1103/PhysRevA.85.032119},
  url = {https://link.aps.org/doi/10.1103/PhysRevA.85.032119}
}

@article{Supic2020selftestingof,
  doi = {10.22331/q-2020-09-30-337},
  url = {https://doi.org/10.22331/q-2020-09-30-337},
  title = {Self-testing of quantum systems: a review},
  author = {{\v{S}}upi{\'{c}}, Ivan and Bowles, Joseph},
  journal = {{Quantum}},
  issn = {2521-327X},
  publisher = {{Verein zur F{\"{o}}rderung des Open Access Publizierens in den Quantenwissenschaften}},
  volume = {4},
  pages = {337},
  month = sep,
  year = {2020}
}

@article{Tavakoli_2022,
doi = {10.1088/1361-6633/ac41bb},
url = {https://dx.doi.org/10.1088/1361-6633/ac41bb},
year = {2022},
month = {mar},
publisher = {IOP Publishing},
volume = {85},
number = {5},
pages = {056001},
author = {Tavakoli, Armin and Pozas-Kerstjens, Alejandro and Luo, Ming-Xing and Renou, Marc-Olivier},
title = {Bell nonlocality in networks},
journal = {Reports on Progress in Physics}
}

@article{sdi_exp_certification_PhysRevLett.113.080405,
  title = {Experimental Semi-Device-Independent Certification of Entangled Measurements},
  author = {Bennet, Adam and V\'ertesi, Tam\'as and Saunders, Dylan J. and Brunner, Nicolas and Pryde, G. J.},
  journal = {Phys. Rev. Lett.},
  volume = {113},
  issue = {8},
  pages = {080405},
  numpages = {5},
  year = {2014},
  month = {Aug},
  publisher = {American Physical Society},
  doi = {10.1103/PhysRevLett.113.080405},
  url = {https://link.aps.org/doi/10.1103/PhysRevLett.113.080405}
}

@article{sdi_certification_2011_PhysRevA.83.062112,
  title = {Certifying entangled measurements in known Hilbert spaces},
  author = {V\'ertesi, Tam\'as and Navascu\'es, Miguel},
  journal = {Phys. Rev. A},
  volume = {83},
  issue = {6},
  pages = {062112},
  numpages = {7},
  year = {2011},
  month = {Jun},
  publisher = {American Physical Society},
  doi = {10.1103/PhysRevA.83.062112},
  url = {https://link.aps.org/doi/10.1103/PhysRevA.83.062112}
}

@article{post_selection_PhysRevLett.107.050502,
  title = {Device-Independent Certification of Entangled Measurements},
  author = {Rabelo, Rafael and Ho, Melvyn and Cavalcanti, Daniel and Brunner, Nicolas and Scarani, Valerio},
  journal = {Phys. Rev. Lett.},
  volume = {107},
  issue = {5},
  pages = {050502},
  numpages = {4},
  year = {2011},
  month = {Jul},
  publisher = {American Physical Society},
  doi = {10.1103/PhysRevLett.107.050502},
  url = {https://link.aps.org/doi/10.1103/PhysRevLett.107.050502}
}

@article{causal_assumptions_in_bell_theorem_PhysRevLett.114.140403,
  title = {Unifying Framework for Relaxations of the Causal Assumptions in Bell's Theorem},
  author = {Chaves, R. and Kueng, R. and Brask, J. B. and Gross, D.},
  journal = {Phys. Rev. Lett.},
  volume = {114},
  issue = {14},
  pages = {140403},
  numpages = {5},
  year = {2015},
  month = {Apr},
  publisher = {American Physical Society},
  doi = {10.1103/PhysRevLett.114.140403},
  url = {https://link.aps.org/doi/10.1103/PhysRevLett.114.140403}
}

@article{Wood_2015,
doi = {10.1088/1367-2630/17/3/033002},
url = {https://dx.doi.org/10.1088/1367-2630/17/3/033002},
year = {2015},
month = {mar},
publisher = {IOP Publishing},
volume = {17},
number = {3},
pages = {033002},
author = {Wood, Christopher J and Spekkens, Robert W},
title = {The lesson of causal discovery algorithms for quantum correlations: causal explanations of Bell-inequality violations require fine-tuning},
journal = {New Journal of Physics}
}

@article{E._G._Cavalcanti_PhysRevX.8.021018,
  title = {Classical Causal Models for Bell and Kochen-Specker Inequality Violations Require Fine-Tuning},
  author = {Cavalcanti, Eric G.},
  journal = {Phys. Rev. X},
  volume = {8},
  issue = {2},
  pages = {021018},
  numpages = {9},
  year = {2018},
  month = {Apr},
  publisher = {American Physical Society},
  doi = {10.1103/PhysRevX.8.021018},
  url = {https://link.aps.org/doi/10.1103/PhysRevX.8.021018}
}

@article{seft_testing_in_prepare_measure_PhysRevA.98.062307,
  title = {Self-testing quantum states and measurements in the prepare-and-measure scenario},
  author = {Tavakoli, Armin and Kaniewski, J\ifmmode \mbox{\k{e}}\else \k{e}\fi{}drzej and V\'ertesi, Tam\'as and Rosset, Denis and Brunner, Nicolas},
  journal = {Phys. Rev. A},
  volume = {98},
  issue = {6},
  pages = {062307},
  numpages = {13},
  year = {2018},
  month = {Dec},
  publisher = {American Physical Society},
  doi = {10.1103/PhysRevA.98.062307},
  url = {https://link.aps.org/doi/10.1103/PhysRevA.98.062307}
}

@article{bell_nonlocality_2014_RevModPhys.86.419,
  title = {Bell nonlocality},
  author = {Brunner, Nicolas and Cavalcanti, Daniel and Pironio, Stefano and Scarani, Valerio and Wehner, Stephanie},
  journal = {Rev. Mod. Phys.},
  volume = {86},
  issue = {2},
  pages = {419--478},
  numpages = {60},
  year = {2014},
  month = {Apr},
  publisher = {American Physical Society},
  doi = {10.1103/RevModPhys.86.419},
  url = {https://link.aps.org/doi/10.1103/RevModPhys.86.419}
}

@article{Quantum_cryptography_2002_RevModPhys.74.145,
  title = {Quantum cryptography},
  author = {Gisin, Nicolas and Ribordy, Gr\'egoire and Tittel, Wolfgang and Zbinden, Hugo},
  journal = {Rev. Mod. Phys.},
  volume = {74},
  issue = {1},
  pages = {145--195},
  numpages = {0},
  year = {2002},
  month = {Mar},
  publisher = {American Physical Society},
  doi = {10.1103/RevModPhys.74.145},
  url = {https://link.aps.org/doi/10.1103/RevModPhys.74.145}
}

@misc{mayers2004selftestingquantumapparatus,
      title={Self testing quantum apparatus}, 
      author={Dominic Mayers and Andrew Yao},
      year={2004},
      eprint={quant-ph/0307205},
      archivePrefix={arXiv},
      primaryClass={quant-ph},
      url={https://arxiv.org/abs/quant-ph/0307205}, 
}

@article{Lee2019,
  author  = {Ciarán M. Lee},
  title   = {Device-independent certification of non-classical joint measurements via causal models},
  journal = {npj Quantum Information},
  year    = {2019},
  volume  = {5},
  number  = {1},
  pages   = {34},
  doi     = {10.1038/s41534-019-0151-1},
  url     = {https://doi.org/10.1038/s41534-019-0151-1},
  issn    = {2056-6387}
}

@misc{ambainis2009quantumrandomaccesscodes,
      title={Quantum Random Access Codes with Shared Randomness}, 
      author={Andris Ambainis and Debbie Leung and Laura Mancinska and Maris Ozols},
      year={2009},
      eprint={0810.2937},
      archivePrefix={arXiv},
      primaryClass={quant-ph},
      url={https://arxiv.org/abs/0810.2937}, 
}

@article{A.K.pan_2022_PhysRevA.105.032216,
  title = {Characterizing nonlocal correlations through various $n$-locality inequalities in a quantum network},
  author = {Munshi, Sneha and Pan, A. K.},
  journal = {Phys. Rev. A},
  volume = {105},
  issue = {3},
  pages = {032216},
  numpages = {11},
  year = {2022},
  month = {Mar},
  publisher = {American Physical Society},
  doi = {10.1103/PhysRevA.105.032216},
  url = {https://link.aps.org/doi/10.1103/PhysRevA.105.032216}
}

@article{sneha_et_al._2023_PhysRevA.108.062607,
  title = {Self-testing of an unbounded number of mutually commuting local observables},
  author = {Munshi, Sneha and Pan, A. K.},
  journal = {Phys. Rev. A},
  volume = {108},
  issue = {6},
  pages = {062607},
  numpages = {11},
  year = {2023},
  month = {Dec},
  publisher = {American Physical Society},
  doi = {10.1103/PhysRevA.108.062607},
  url = {https://link.aps.org/doi/10.1103/PhysRevA.108.062607}
}

@article{Ghorai_et_al.PhysRevA.98.032110,
  title = {Optimal quantum preparation contextuality in an $n$-bit parity-oblivious multiplexing task},
  author = {Ghorai, Shouvik and Pan, A. K.},
  journal = {Phys. Rev. A},
  volume = {98},
  issue = {3},
  pages = {032110},
  numpages = {8},
  year = {2018},
  month = {Sep},
  publisher = {American Physical Society},
  doi = {10.1103/PhysRevA.98.032110},
  url = {https://link.aps.org/doi/10.1103/PhysRevA.98.032110}
}

@article{armin_2024_non_projective_BSM_PhysRevA.110.042206,
  title = {Nonprojective Bell-state measurements},
  author = {Wei, Amanda and Cobucci, Gabriele and Tavakoli, Armin},
  journal = {Phys. Rev. A},
  volume = {110},
  issue = {4},
  pages = {042206},
  numpages = {8},
  year = {2024},
  month = {Oct},
  publisher = {American Physical Society},
  doi = {10.1103/PhysRevA.110.042206},
  url = {https://link.aps.org/doi/10.1103/PhysRevA.110.042206}
}

@article{Armin_Tavakoli_et_al._Self-testing_nonprojective_quantum_measurements_doi:10.1126/sciadv.aaw6664,
author = {Armin Tavakoli  and Massimiliano Smania  and Tamás Vértesi  and Nicolas Brunner  and Mohamed Bourennane },
title = {Self-testing nonprojective quantum measurements in prepare-and-measure experiments},
journal = {Science Advances},
volume = {6},
number = {16},
pages = {eaaw6664},
year = {2020},
doi = {10.1126/sciadv.aaw6664},
URL = {https://www.science.org/doi/abs/10.1126/sciadv.aaw6664},
eprint = {https://www.science.org/doi/pdf/10.1126/sciadv.aaw6664}}

@article{Sarkar2026,
  author  = {Sarkar, Shubhayan and Orthey Jr, Alexandre C. and Augusiak, Remigiusz},
  title   = {A universal scheme to self-test any quantum state or measurement},
  journal = {Nature Physics},
  volume  = {22},
  number  = {3},
  pages   = {446--451},
  year    = {2026},
  doi     = {10.1038/s41567-026-03181-y},
  url     = {https://doi.org/10.1038/s41567-026-03181-y}
}

@article{supic_2020_PM,
  author  = {Šupić, Ivan and Hoban, Matty J. and Colomer, Laia Domingo and Acín, Antonio},
  title   = {Self-testing and certification using trusted quantum inputs},
  journal = {New Journal of Physics},
  volume  = {22},
  number  = {7},
  pages   = {073006},
  year    = {2020},
  month   = {jul},
  publisher = {IOP Publishing},
  doi     = {10.1088/1367-2630/ab90d1},
  url     = {https://doi.org/10.1088/1367-2630/ab90d1}
}

@article{pauwels_2022_PM_PhysRevLett.129.250504,
  title = {Almost Qudits in the Prepare-and-Measure Scenario},
  author = {Pauwels, Jef and Pironio, Stefano and Woodhead, Erik and Tavakoli, Armin},
  journal = {Phys. Rev. Lett.},
  volume = {129},
  issue = {25},
  pages = {250504},
  numpages = {7},
  year = {2022},
  month = {Dec},
  publisher = {American Physical Society},
  doi = {10.1103/PhysRevLett.129.250504},
  url = {https://link.aps.org/doi/10.1103/PhysRevLett.129.250504}
}

@article{Navascués_2023_PM_PhysRevLett.131.250802,
  title = {Self-Testing in Prepare-and-Measure Scenarios and a Robust Version of Wigner's Theorem},
  author = {Navascu\'es, Miguel and P\'al, K\'aroly F. and V\'ertesi, Tam\'as and Ara\'ujo, Mateus},
  journal = {Phys. Rev. Lett.},
  volume = {131},
  issue = {25},
  pages = {250802},
  numpages = {6},
  year = {2023},
  month = {Dec},
  publisher = {American Physical Society},
  doi = {10.1103/PhysRevLett.131.250802},
  url = {https://link.aps.org/doi/10.1103/PhysRevLett.131.250802}
}

@article{Debashis_2026_GHZ_mvmn-fzq7,
  title = {Communication scenario enabling semi-device-independent robust self-testing of $n$-party Greenberger-Horne-Zeilinger basis measurements},
  author = {Bhaumik, Barnik and Ray, Sagnik and Saha, Debashis},
  journal = {Phys. Rev. A},
  volume = {113},
  issue = {1},
  pages = {012407},
  numpages = {17},
  year = {2026},
  month = {Jan},
  publisher = {American Physical Society},
  doi = {10.1103/mvmn-fzq7},
  url = {https://link.aps.org/doi/10.1103/mvmn-fzq7}
}

@article{Dense_coding_1992,
  title = {Communication via one- and two-particle operators on Einstein-Podolsky-Rosen states},
  author = {Bennett, Charles H. and Wiesner, Stephen J.},
  journal = {Phys. Rev. Lett.},
  volume = {69},
  issue = {20},
  pages = {2881--2884},
  numpages = {0},
  year = {1992},
  month = {Nov},
  publisher = {American Physical Society},
  doi = {10.1103/PhysRevLett.69.2881},
  url = {https://link.aps.org/doi/10.1103/PhysRevLett.69.2881}
}

@article{Teleportation_95,
  title = {Teleporting an unknown quantum state via dual classical and Einstein-Podolsky-Rosen channels},
  author = {Bennett, Charles H. and Brassard, Gilles and Cr\'epeau, Claude and Jozsa, Richard and Peres, Asher and Wootters, William K.},
  journal = {Phys. Rev. Lett.},
  volume = {70},
  issue = {13},
  pages = {1895--1899},
  numpages = {0},
  year = {1993},
  month = {Mar},
  publisher = {American Physical Society},
  doi = {10.1103/PhysRevLett.70.1895},
  url = {https://link.aps.org/doi/10.1103/PhysRevLett.70.1895}
}

@article{Renou_2018,
   title={Self-Testing Entangled Measurements in Quantum Networks},
   volume={121},
   ISSN={1079-7114},
   url={http://dx.doi.org/10.1103/PhysRevLett.121.250507},
   DOI={10.1103/physrevlett.121.250507},
   number={25},
   journal={Physical Review Letters},
   publisher={American Physical Society (APS)},
   author={Renou, Marc Olivier and Kaniewski, Jędrzej and Brunner, Nicolas},
   year={2018},
   month=Dec }

@article{Frenkel_2015,
   title={Classical Information Storage in an n-Level Quantum System},
   volume={340},
   ISSN={1432-0916},
   url={http://dx.doi.org/10.1007/s00220-015-2463-0},
   DOI={10.1007/s00220-015-2463-0},
   number={2},
   journal={Communications in Mathematical Physics},
   publisher={Springer Science and Business Media LLC},
   author={Frenkel, Péter E. and Weiner, Mihály},
   year={2015},
   month=Sept, pages={563–574} }

@Article{entropy_21030325,
AUTHOR = {Gisin, Nicolas},
TITLE = {Genuine Quantum Nonlocality in the: Testing Joint Measurements in Quantum Networks},
JOURNAL = {Entropy},
VOLUME = {21},
YEAR = {2019},
NUMBER = {3},
ARTICLE-NUMBER = {325},
URL = {https://www.mdpi.com/1099-4300/21/3/325},
PubMedID = {33267039},
ISSN = {1099-4300},
DOI = {10.3390/e21030325}
}

@article{Bilocal_EJM_PhysRevLett.126.220401,
  title = {Bilocal Bell Inequalities Violated by the Quantum Elegant Joint Measurement},
  author = {Tavakoli, Armin and Gisin, Nicolas and Branciard, Cyril},
  journal = {Phys. Rev. Lett.},
  volume = {126},
  issue = {22},
  pages = {220401},
  numpages = {6},
  year = {2021},
  month = {Jun},
  publisher = {American Physical Society},
  doi = {10.1103/PhysRevLett.126.220401},
  url = {https://link.aps.org/doi/10.1103/PhysRevLett.126.220401}
}

@article{ritesh_Singh_2025,
doi = {10.1088/1367-2630/ae1f35},
url = {https://doi.org/10.1088/1367-2630/ae1f35},
year = {2025},
month = {nov},
publisher = {IOP Publishing},
volume = {27},
number = {11},
pages = {114520},
author = {Singh, Ritesh K and Sasmal, Souradeep and Nautiyal, S and Pan, A K},
title = {Self-testing in a constrained prepare-measure scenario sans assuming quantum dimension},
journal = {New Journal of Physics}
}

@article{Paul_2025,
   title={Semi-device-independent self-testing of unitary operations},
   volume={112},
   ISSN={2469-9934},
   url={http://dx.doi.org/10.1103/94pf-njhr},
   DOI={10.1103/94pf-njhr},
   number={6},
   journal={Physical Review A},
   publisher={American Physical Society (APS)},
   author={Paul, Rajdeep and Roy, Prabuddha and Pan, A. K.},
   year={2025},
   month=Dec }

@article{ritesh_GHz_6ct-vmnx,
  title = {Self-testing of the $m$-partite Greenberger-Horne-Zeilinger state and observables using the Svetlichny inequality},
  author = {Singh, Ritesh K. and Sasmal, Souradeep and Pan, A. K.},
  journal = {Phys. Rev. A},
  volume = {112},
  issue = {3},
  pages = {032404},
  numpages = {30},
  year = {2025},
  month = {Sep},
  publisher = {American Physical Society},
  doi = {10.1103/x6ct-vmnx},
  url = {https://link.aps.org/doi/10.1103/x6ct-vmnx}
}

@misc{ambainis1998densequantumcodinglower,
      title={Dense Quantum Coding and a Lower Bound for 1-way Quantum Automata}, 
      author={Andris Ambainis and Ashwin Nayak and Amnon Ta-Shma and Umesh Vazirani},
      year={1998},
      eprint={quant-ph/9804043},
      archivePrefix={arXiv},
      primaryClass={quant-ph},
      url={https://arxiv.org/abs/quant-ph/9804043}, 
}

@article{Holevo1973,
  author  = {Holevo, A. S.},
  title   = {Bounds for the Quantity of Information Transmitted
             by a Quantum Communication Channel},
  journal = {Problems of Information Transmission},
  year    = {1973},
  volume  = {9},
  number  = {3},
  pages   = {177--183},
  note    = {English translation}
}

@article{Rahul_2021_PhysRevA.104.042217,
  title = {Generalized $n$-locality inequalities in a star-network configuration and their optimal quantum violations},
  author = {Munshi, Sneha and Kumar, Rahul and Pan, A. K.},
  journal = {Phys. Rev. A},
  volume = {104},
  issue = {4},
  pages = {042217},
  numpages = {10},
  year = {2021},
  month = {Oct},
  publisher = {American Physical Society},
  doi = {10.1103/PhysRevA.104.042217},
  url = {https://link.aps.org/doi/10.1103/PhysRevA.104.042217}
}

@article{Tavakoli_2018,
   title={Semi-device-independent characterization of multipartite entanglement of states and measurements},
   volume={98},
   ISSN={2469-9934},
   url={http://dx.doi.org/10.1103/PhysRevA.98.052333},
   DOI={10.1103/physreva.98.052333},
   number={5},
   journal={Physical Review A},
   publisher={American Physical Society (APS)},
   author={Tavakoli, Armin and Abbott, Alastair A. and Renou, Marc-Olivier and Gisin, Nicolas and Brunner, Nicolas},
   year={2018},
   month=Nov }

@article{Brunner_et.al_SDIQKD_2011_PhysRevA.84.010302,
  title = {Semi-device-independent security of one-way quantum key distribution},
  author = {Paw{\l{}}owski, Marcin and Brunner, Nicolas},
  journal = {Phys. Rev. A},
  volume = {84},
  issue = {1},
  pages = {010302(R)},
  numpages = {4},
  year = {2011},
  month = {Jul},
  publisher = {American Physical Society},
  doi = {10.1103/PhysRevA.84.010302},
  url = {https://link.aps.org/doi/10.1103/PhysRevA.84.010302}
}

@article{_Tavakoli_et.al_QRAC_PhysRevLett.114.170502,
  title = {Quantum Random Access Codes Using Single $d$-Level Systems},
  author = {Tavakoli, Armin and Hameedi, Alley and Marques, Breno and Bourennane, Mohamed},
  journal = {Phys. Rev. Lett.},
  volume = {114},
  issue = {17},
  pages = {170502},
  numpages = {5},
  year = {2015},
  month = {Apr},
  publisher = {American Physical Society},
  doi = {10.1103/PhysRevLett.114.170502},
  url = {https://link.aps.org/doi/10.1103/PhysRevLett.114.170502}
}

@article{vertesi_et.al_dimension_witness_2013_PhysRevLett.110.150501,
  title = {Dimension Witnesses and Quantum State Discrimination},
  author = {Brunner, Nicolas and Navascu{\'e}s, Miguel and V{\'e}rtesi, Tam{\'a}s},
  journal = {Phys. Rev. Lett.},
  volume = {110},
  issue = {15},
  pages = {150501},
  numpages = {4},
  year = {2013},
  month = {Apr},
  publisher = {American Physical Society},
  doi = {10.1103/PhysRevLett.110.150501},
  url = {https://link.aps.org/doi/10.1103/PhysRevLett.110.150501}
}

@article{VanHimbeeck2017semidevice,
  doi = {10.22331/q-2017-11-18-33},
  url = {https://doi.org/10.22331/q-2017-11-18-33},
  title = {Semi-device-independent framework based on natural physical assumptions},
  author = {Van Himbeeck, Thomas and Woodhead, Erik and Cerf, Nicolas J. and Garc{\'{i}}a-Patr{\'{o}}n, Ra{\'{u}}l and Pironio, Stefano},
  journal = {{Quantum}},
  issn = {2521-327X},
  publisher = {{Verein zur F{\"{o}}rderung des Open Access Publizierens in den Quantenwissenschaften}},
  volume = {1},
  pages = {33},
  month = nov,
  year = {2017}
}

@article{Chailloux_2016,
doi = {10.1088/1367-2630/18/4/045003},
url = {https://doi.org/10.1088/1367-2630/18/4/045003},
year = {2016},
month = {apr},
publisher = {IOP Publishing},
volume = {18},
number = {4},
pages = {045003},
author = {Chailloux, André and Kerenidis, Iordanis and Kundu, Srijita and Sikora, Jamie},
title = {Optimal bounds for parity-oblivious random access codes},
journal = {New Journal of Physics}
}

@article{Mohan_2019,
doi = {10.1088/1367-2630/ab3773},
url = {https://doi.org/10.1088/1367-2630/ab3773},
year = {2019},
month = {aug},
publisher = {IOP Publishing},
volume = {21},
number = {8},
pages = {083034},
author = {Mohan, Karthik and Tavakoli, Armin and Brunner, Nicolas},
title = {Sequential random access codes and self-testing of quantum measurement instruments},
journal = {New Journal of Physics}
}

@article{Sumit_2021PhysRevA.104.062214,
  title = {Semi-device-independent certification of multiple unsharpness parameters through sequential measurements},
  author = {Mukherjee, Sumit and Pan, A. K.},
  journal = {Phys. Rev. A},
  volume = {104},
  issue = {6},
  pages = {062214},
  numpages = {16},
  year = {2021},
  month = {Dec},
  publisher = {American Physical Society},
  doi = {10.1103/PhysRevA.104.062214},
  url = {https://link.aps.org/doi/10.1103/PhysRevA.104.062214}
}

@article{Wei_2021,
doi = {10.1088/1367-2630/abf614},
url = {https://doi.org/10.1088/1367-2630/abf614},
year = {2021},
month = {may},
publisher = {IOP Publishing},
volume = {23},
number = {5},
pages = {053014},
author = {Wei, Shihui and Guo, Fenzhuo and Gao, Fei and Wen, Qiaoyan},
title = {Certification of three black boxes with unsharp measurements using 3 → 1 sequential quantum random access codes},
journal = {New Journal of Physics}
}

@article{Abhyoudai_et.al_PhysRevA.107.012411,
  title = {Robust certification of unsharp instruments through sequential quantum advantages in a prepare-measure communication game},
  author = {S. S., Abhyoudai and Mukherjee, Sumit and Pan, A. K.},
  journal = {Phys. Rev. A},
  volume = {107},
  issue = {1},
  pages = {012411},
  numpages = {18},
  year = {2023},
  month = {Jan},
  publisher = {American Physical Society},
  doi = {10.1103/PhysRevA.107.012411},
  url = {https://link.aps.org/doi/10.1103/PhysRevA.107.012411}
}

@article{Tavakoli_fidelity_PhysRevA.100.042108,
  title = {Semi-device-independent characterization of quantum measurements under a minimum overlap assumption},
  author = {Shi, Weixu and Cai, Yu and Brask, Jonatan Bohr and Zbinden, Hugo and Brunner, Nicolas},
  journal = {Phys. Rev. A},
  volume = {100},
  issue = {4},
  pages = {042108},
  numpages = {9},
  year = {2019},
  month = {Oct},
  publisher = {American Physical Society},
  doi = {10.1103/PhysRevA.100.042108},
  url = {https://link.aps.org/doi/10.1103/PhysRevA.100.042108}
}

@article{Burnner_etal_overlap_PhysRevApplied.7.054018,
  title = {Megahertz-Rate Semi-Device-Independent Quantum Random Number Generators Based on Unambiguous State Discrimination},
  author = {Brask, Jonatan Bohr and Martin, Anthony and Esposito, William and Houlmann, Raphael and Bowles, Joseph and Zbinden, Hugo and Brunner, Nicolas},
  journal = {Phys. Rev. Appl.},
  volume = {7},
  issue = {5},
  pages = {054018},
  numpages = {10},
  year = {2017},
  month = {May},
  publisher = {American Physical Society},
  doi = {10.1103/PhysRevApplied.7.054018},
  url = {https://link.aps.org/doi/10.1103/PhysRevApplied.7.054018}
}

@article{Tavakoli2020informationally,
  doi = {10.22331/q-2020-09-24-332},
  url = {https://doi.org/10.22331/q-2020-09-24-332},
  title = {Informationally restricted quantum correlations},
  author = {Tavakoli, Armin and Zambrini Cruzeiro, Emmanuel and Bohr Brask, Jonatan and Gisin, Nicolas and Brunner, Nicolas},
  journal = {{Quantum}},
  issn = {2521-327X},
  publisher = {{Verein zur F{\"{o}}rderung des Open Access Publizierens in den Quantenwissenschaften}},
  volume = {4},
  pages = {332},
  month = sep,
  year = {2020}
}

@article{Chaves_2015_PhysRevLett.115.110501,
  title = {Device-Independent Tests of Entropy},
  author = {Chaves, Rafael and Brask, Jonatan Bohr and Brunner, Nicolas},
  journal = {Phys. Rev. Lett.},
  volume = {115},
  issue = {11},
  pages = {110501},
  numpages = {5},
  year = {2015},
  month = {Sep},
  publisher = {American Physical Society},
  doi = {10.1103/PhysRevLett.115.110501},
  url = {https://link.aps.org/doi/10.1103/PhysRevLett.115.110501}
}

@article{Tavakoli2022informationally,
  doi = {10.22331/q-2022-01-05-620},
  url = {https://doi.org/10.22331/q-2022-01-05-620},
  title = {Informationally restricted correlations: a general framework for classical and quantum systems},
  author = {Tavakoli, Armin and Zambrini Cruzeiro, Emmanuel and Woodhead, Erik and Pironio, Stefano},
  journal = {{Quantum}},
  issn = {2521-327X},
  publisher = {{Verein zur F{\"{o}}rderung des Open Access Publizierens in den Quantenwissenschaften}},
  volume = {6},
  pages = {620},
  month = jan,
  year = {2022}
}

@article{Supic2023QuantumNetworksSelfTest,
  author  = {Ivan {\v S}upi{\'c} and Joseph Bowles and Marc-Olivier Renou and Antonio Ac{\'i}n and Matty J. Hoban},
  title   = {Quantum networks self-test all entangled states},
  journal = {Nature Physics},
  year    = {2023},
  month   = may,
  volume  = {19},
  number  = {5},
  pages   = {670--675},
  issn    = {1745-2481},
  doi     = {10.1038/s41567-023-01945-4},
  url     = {https://doi.org/10.1038/s41567-023-01945-4}
}

@article{Sarkar2021,
  author    = {Sarkar, Shubhayan and Saha, Debashis and Kaniewski, J{\k e}drzej and Augusiak, Remigiusz},
  title     = {Self-testing quantum systems of arbitrary local dimension with minimal number of measurements},
  journal   = {npj Quantum Information},
  year      = {2021},
  volume    = {7},
  number    = {1},
  pages     = {151},
  doi       = {10.1038/s41534-021-00490-3},
  url       = {https://doi.org/10.1038/s41534-021-00490-3},
  issn      = {2056-6387}
}

@article{Wu2014,
  title = {Robust self-testing of the three-qubit $W$ state},
  author = {Wu, Xingyao and Cai, Yu and Yang, Tzyh Haur and Le, Huy Nguyen and Bancal, Jean-Daniel and Scarani, Valerio},
  journal = {Phys. Rev. A},
  volume = {90},
  issue = {4},
  pages = {042339},
  numpages = {10},
  year = {2014},
  month = {Oct},
  publisher = {American Physical Society},
  doi = {10.1103/PhysRevA.90.042339},
  url = {https://link.aps.org/doi/10.1103/PhysRevA.90.042339}
}

@article{Panwar2023,
  author    = {Panwar, Ekta and Pandya, Palash and Wie{\'s}niak, Marcin},
  title     = {An elegant scheme of self-testing for multipartite Bell inequalities},
  journal   = {npj Quantum Information},
  year      = {2023},
  volume    = {9},
  number    = {1},
  pages     = {71},
  doi       = {10.1038/s41534-023-00735-3},
  url       = {https://doi.org/10.1038/s41534-023-00735-3},
  issn      = {2056-6387}
}

@InProceedings{McKague10.1007/978-3-642-54429-3_7,
author="McKague, Matthew",
editor="Bacon, Dave
and Martin-Delgado, Miguel
and Roetteler, Martin",
title="Self-Testing Graph States",
booktitle="Theory of Quantum Computation, Communication, and Cryptography",
year="2014",
publisher="Springer Berlin Heidelberg",
address="Berlin, Heidelberg",
pages="104--120",
isbn="978-3-642-54429-3"
}

@article{Acin2012,
  title = {Randomness versus Nonlocality and Entanglement},
  author = {Ac{\'{\i}}n, Antonio and Massar, Serge and Pironio, Stefano},
  journal = {Phys. Rev. Lett.},
  volume = {108},
  issue = {10},
  pages = {100402},
  numpages = {5},
  year = {2012},
  month = {Mar},
  publisher = {American Physical Society},
  doi = {10.1103/PhysRevLett.108.100402},
  url = {https://link.aps.org/doi/10.1103/PhysRevLett.108.100402}
}

@article{Coladangelo2017,
  author    = {Coladangelo, Andrea and Goh, Koon Tong and Scarani, Valerio},
  title     = {All pure bipartite entangled states can be self-tested},
  journal   = {Nature Communications},
  year      = {2017},
  volume    = {8},
  number    = {1},
  pages     = {15485},
  doi       = {10.1038/ncomms15485},
  url       = {https://doi.org/10.1038/ncomms15485},
  issn      = {2041-1723}
}

@article{Wooltron2022,
  title = {Tight Analytic Bound on the Trade-Off between Device-Independent Randomness and Nonlocality},
  author = {Wooltorton, Lewis and Brown, Peter and Colbeck, Roger},
  journal = {Phys. Rev. Lett.},
  volume = {129},
  issue = {15},
  pages = {150403},
  numpages = {6},
  year = {2022},
  month = {Oct},
  publisher = {American Physical Society},
  doi = {10.1103/PhysRevLett.129.150403},
  url = {https://link.aps.org/doi/10.1103/PhysRevLett.129.150403}
}

@article{Rai2022,
  title = {Self-testing quantum states via nonmaximal violation in Hardy's test of nonlocality},
  author = {Rai, Ashutosh and Pivoluska, Matej and Sasmal, Souradeep and Banik, Manik and Ghosh, Sibasish and Plesch, Martin},
  journal = {Phys. Rev. A},
  volume = {105},
  issue = {5},
  pages = {052227},
  numpages = {8},
  year = {2022},
  month = {May},
  publisher = {American Physical Society},
  doi = {10.1103/PhysRevA.105.052227},
  url = {https://link.aps.org/doi/10.1103/PhysRevA.105.052227}
}

@article{Bamps2015,
  title = {Sum-of-squares decompositions for a family of Clauser-Horne-Shimony-Holt-like inequalities and their application to self-testing},
  author = {Bamps, C{\'e}dric and Pironio, Stefano},
  journal = {Phys. Rev. A},
  volume = {91},
  issue = {5},
  pages = {052111},
  numpages = {13},
  year = {2015},
  month = {May},
  publisher = {American Physical Society},
  doi = {10.1103/PhysRevA.91.052111},
  url = {https://link.aps.org/doi/10.1103/PhysRevA.91.052111}
}

@article{Bowles2018,
  title = {Self-testing of Pauli observables for device-independent entanglement certification},
  author = {Bowles, Joseph and {\ifmmode \check{S}\else \v{S}\fi{}}upi{\ifmmode \acute{c}\else \'{c}\fi{}}, Ivan and Cavalcanti, Daniel and Ac{\'{\i}}n, Antonio},
  journal = {Phys. Rev. A},
  volume = {98},
  issue = {4},
  pages = {042336},
  numpages = {24},
  year = {2018},
  month = {Oct},
  publisher = {American Physical Society},
  doi = {10.1103/PhysRevA.98.042336},
  url = {https://link.aps.org/doi/10.1103/PhysRevA.98.042336}
}

@article{Pan2021,
  title = {Oblivious communication game, self-testing of projective and nonprojective measurements, and certification of randomness},
  author = {Pan, A. K.},
  journal = {Phys. Rev. A},
  volume = {104},
  issue = {2},
  pages = {022212},
  numpages = {14},
  year = {2021},
  month = {Aug},
  publisher = {American Physical Society},
  doi = {10.1103/PhysRevA.104.022212},
  url = {https://link.aps.org/doi/10.1103/PhysRevA.104.022212}
}

@article{Wagner2020,
  doi = {10.22331/q-2020-03-19-243},
  url = {https://doi.org/10.22331/q-2020-03-19-243},
  title = {Device-independent characterization of quantum instruments},
  author = {Wagner, Sebastian and Bancal, Jean-Daniel and Sangouard, Nicolas and Sekatski, Pavel},
  journal = {{Quantum}},
  issn = {2521-327X},
  publisher = {{Verein zur F{\"{o}}rderung des Open Access Publizierens in den Quantenwissenschaften}},
  volume = {4},
  pages = {243},
  month = mar,
  year = {2020}
}

@article{SekatskiPrl2018,
  title = {Certifying the Building Blocks of Quantum Computers from Bell's Theorem},
  author = {Sekatski, Pavel and Bancal, Jean-Daniel and Wagner, Sebastian and Sangouard, Nicolas},
  journal = {Phys. Rev. Lett.},
  volume = {121},
  issue = {18},
  pages = {180505},
  numpages = {5},
  year = {2018},
  month = {Nov},
  publisher = {American Physical Society},
  doi = {10.1103/PhysRevLett.121.180505},
  url = {https://link.aps.org/doi/10.1103/PhysRevLett.121.180505}
}

@article{SekatskiPrl2023,
  title = {Toward the Device-Independent Certification of a Quantum Memory},
  author = {Sekatski, Pavel and Bancal, Jean-Daniel and Ioannou, Marie and Afzelius, Mikael and Brunner, Nicolas},
  journal = {Phys. Rev. Lett.},
  volume = {131},
  issue = {17},
  pages = {170802},
  numpages = {6},
  year = {2023},
  month = {Oct},
  publisher = {American Physical Society},
  doi = {10.1103/PhysRevLett.131.170802},
  url = {https://link.aps.org/doi/10.1103/PhysRevLett.131.170802}
}

@article{Roy2023,
doi = {10.1088/1367-2630/acb4b5},
url = {https://doi.org/10.1088/1367-2630/acb4b5},
year = {2023},
month = {feb},
publisher = {IOP Publishing},
volume = {25},
number = {1},
pages = {013040},
author = {Roy, Prabuddha and Pan, A K},
title = {Device-independent self-testing of unsharp measurements},
journal = {New Journal of Physics}
}

@article{Paul2024,
  title = {Self-testing of multiple unsharpness parameters through sequential violations of a noncontextual inequality},
  author = {Paul, Rajdeep and Sasmal, Souradeep and Pan, A. K.},
  journal = {Phys. Rev. A},
  volume = {110},
  issue = {1},
  pages = {012444},
  numpages = {21},
  year = {2024},
  month = {Jul},
  publisher = {American Physical Society},
  doi = {10.1103/PhysRevA.110.012444},
  url = {https://link.aps.org/doi/10.1103/PhysRevA.110.012444}
}

@article{Gomez2016,
  title = {Device-Independent Certification of a Nonprojective Qubit Measurement},
  author = {G{\'o}mez, Esteban S. and G{\'o}mez, Santiago and Gonz{\'a}lez, Pablo and Ca{\~n}as, Gustavo and Barra, Johanna F. and Delgado, Aldo and Xavier, Guilherme B. and Cabello, Ad{\'a}n and Kleinmann, Matthias and V{\'e}rtesi, Tam{\'a}s and Lima, Gustavo},
  journal = {Phys. Rev. Lett.},
  volume = {117},
  issue = {26},
  pages = {260401},
  numpages = {5},
  year = {2016},
  month = {Dec},
  publisher = {American Physical Society},
  doi = {10.1103/PhysRevLett.117.260401},
  url = {https://link.aps.org/doi/10.1103/PhysRevLett.117.260401}
}

@article{Renou2021,
  author    = {Renou, Marc-Olivier and Trillo, David and Weilenmann, Mirjam and Le, Thinh P. and Tavakoli, Armin and Gisin, Nicolas and Ac{\'i}n, Antonio and Navascu{\'e}s, Miguel},
  title     = {Quantum theory based on real numbers can be experimentally falsified},
  journal   = {Nature},
  year      = {2021},
  volume    = {600},
  number    = {7890},
  pages     = {625--629},
  doi       = {10.1038/s41586-021-04160-4},
  url       = {https://doi.org/10.1038/s41586-021-04160-4},
  issn      = {1476-4687}
}
\end{document}